\documentclass[prc,showpacs,preprintnumbers,amsmath,amssymb,aps,nofootinbib,floatfix]{revtex4-1}
\usepackage[utf8]{inputenc}
\usepackage{amsmath}
\usepackage{graphicx}
\usepackage{amsfonts}
\usepackage{amssymb}
\allowdisplaybreaks[1]
\graphicspath{{pics/}}

\begin{document}
\title{Regge spectra of excited mesons, harmonic confinement and QCD vacuum structure}

\author{ Sergei N.  Nedelko\footnote{nedelko@theor.jinr.ru}, Vladimir E. Voronin\footnote{voronin@theor.jinr.ru}}
\affiliation{ Bogoliubov Laboratory of Theoretical Physics, JINR,
141980 Dubna, Russia }
\begin{abstract}

An approach to QCD vacuum as a medium describable in terms of statistical ensemble of almost everywhere  homogeneous Abelian (anti-)self-dual  gluon fields is briefly reviewed.  These fields play the role of the confining medium for color charged fields as well as underline the mechanism of realization of chiral $SU_{\rm L}(N_f)\times SU_{\rm R}(N_f)$  and $U_A(1)$ symmetries.  Hadronization formalism based on this ensemble leads to manifestly defined quantum effective meson action.  Strong, electromagnetic and weak interactions of mesons are represented in the action in terms of nonlocal $n$-point interaction  vertices given by the quark-gluon loops averaged over the background ensemble.  New systematic results for the mass spectrum and decay constants of radially excited  light, heavy-light mesons and heavy quarkonia are presented. Interrelation between  the present approach,   models based on  ideas of soft wall AdS/QCD, light front  holographic QCD,  and  the picture of harmonic confinement is outlined.

\end{abstract}


\maketitle
\section{Introduction}

Almost forty five years ago   Feynman,  Kislinger, and Ravndal noticed~\cite{Feynman} that the Regge spectrum of meson and baryon masses could be universally described  by assuming the four-dimensional harmonic oscillator potential acting between quarks and antiquarks.   During subsequent  years the idea of four-dimensional harmonic oscillator  re-entered  the  discussion about quark confinement  several times  in  various ways. Leutwyler and Stern  developed  the formalism devoted to the  covariant description of bilocal meson-like fields $\Phi(x,z)$ combined with the idea of harmonic confinement~\cite{Leutwyler:1977vz,Leutwyler:1977vy, Leutwyler:1977pv, Leutwyler:1977cs, Leutwyler:1978uk}. Considerations of paper~\cite{Feynman} and Leutwyler-Stern formalism~\cite{Leutwyler:1977vz,Leutwyler:1977vy, Leutwyler:1977pv, Leutwyler:1977cs, Leutwyler:1978uk} can be seen as the forerunners to 
at present time very popular soft wall AdS/QCD models~\cite{Karch:2006pv} and the light front holographic QCD~\cite{deTeramond:2005su,Brodsky:2006uqa,deTeramond:2008ht}.  In recent years,  the approaches to confinement based on the ideas of  soft wall AdS/QCD model and light front holography demonstrated an impressive phenomenological success~\cite{Karch:2006pv,deTeramond:2005su,Brodsky:2006uqa,deTeramond:2008ht,Swarnkar,Gutsche:2015xva}. The crucial  for phenomenology features of these approaches are the particular dilaton profile $\varphi(z)=\kappa^2 z^2$ and the harmonic oscillator form of the confining potential as the function of  fifth coordinate $z$.   All these approaches begin with different motivation but finally come to the Schr\"odinger type differential equation with the harmonic potential in $z$  defining the wave functions and mass spectrum of mesons and baryons.

The physical origin of the above-mentioned particular form of dilaton profile in AdS/QCD and light front holography as well as the harmonic potential in the Stern-Leutwyler studies and, hence,  the Laguerre polynomial form of the meson wave functions, could  not be identified within these approaches  themselves. 
The  preferable form of the dilaton profile and/or the potential are determined by the phenomenological requirement of Regge character of the excited meson  mass spectrum~\cite{Karch:2006pv,Feynman,Leutwyler:1977vz,Leutwyler:1977vy, Leutwyler:1977pv, Leutwyler:1977cs, Leutwyler:1978uk}.

The approach presented in this paper  has been  developed in essence twenty years ago~\cite{EN1}. It   clearly 
incorporates the idea of harmonic confinement both in terms of elementary color charged fields and the composite colorless  hadron field. The distinctive feature of the present approach is that it basically links the concept of harmonic confinement  and Regge character of hadron mass spectrum to the specific class of nonperturbative gluon configurations -- almost everywhere homogeneous Abelian (anti-)self-dual gluon fields. 
A posteriori  a close interrelation of the   Abelian (anti-)self-dual fields and the hadronization based on harmonic confinement   can be read off the  papers~\cite{Leutwyler1, Leutwyler2, Minkowski,  Leutwyler:1977vz,Leutwyler:1977vy, Leutwyler:1977pv, Leutwyler:1977cs, Leutwyler:1978uk}. 
In brief,  the line of arguments is as follows (for more detailed exposition see~\cite{NV}). 

An important benchmark has been the  observation of  Pagels and Tomboulis~\cite{Pagels} that Abelian self-dual fields describe a  medium  infinitely stiff to small gauge field fluctuations, i.e. the  wave solutions for the effective quantum equations of motion are absent. This feature was interpreted as suggestive of confinement of color. 
Strong argumentation in favour of the Abelian (anti-)self-dual homogeneous field as a candidate for the global nontrivial minimum of the effective action originates from the papers \cite{Minkowski,Leutwyler2,NG2011,Pawlowski,George:2012sb}.  In particular, Leutwyler has shown  that the constant gauge field is stable  (tachyon free) against small quantum fluctuations only if it is Abelian (anti-)self-dual covariantly constant field  \cite{Minkowski,Leutwyler2}. Nonperturbative calculation of the effective potential within the functional renormalization group \cite{Pawlowski} supported the earlier one-loop results on existence of the nontrivial minimum of the effective action for the Abelian (anti-)self-dual field. 

The eigenvalues of the Dirac and Klein-Gordon operators in the presence of Abelian self-dual field are purely discrete, and the corresponding eigenfunctions  of quarks and gluons  are of the bound state type. This is a consequence of the 
fact that these operators contain the four-dimensional harmonic oscillator, acting as a confining harmonic potential.  Eigenmodes  of the color charged fields have no (quasi-)particle interpretation but describe field fluctuations decaying in space and time. The consequence of this property is  that the momentum representation of the translation invariant part of the propagator of the color charged field in the background of (anti-)self-dual Abelian gauge field is entire analytical function. 
The absence of pole in the propagator was treated as the absence of the particle interpretation of the charged field~\cite{Leutwyler1}. 
 However just the absence of a single quark or anti-quark in the spectrum can not be considered as  sufficient condition for confinement.  
One has to explain the most peculiar feature of QCD -- the Regge character of the physical spectrum of colorless hadrons. Usually  Regge spectrum  is related to the string picture of confinement, justified in two complementary ways and limits: classical relativistic rotating string connecting massless quark and antiquark, and  the linear potential between nonrelativistic heavy quark and antiquark with the area law for the temporal Wilson loop as a relevant criterion for static quark confinement. Neither the homogeneous Abelian (anti-)self-dual
field itself nor  the form of gluon propagator in the presence of this background
had the clue to linear quark-antiquark potential. Nevertheless, the analytic structure of the gluon and quark propagators and assumption about the randomness of the background field ensemble led both to the area law for static quarks and the Regge spectrum for light hadrons.

 Randomness of the ensemble of almost everywhere homogeneous Abelian (anti-)self-dual gluon fields has been taken into account implicitly in the model of hadronization developed in~\cite{EN1,EN} \textit{via } averaging of the quark loops over the parameters of the random fields. The nonlocal quark-meson vertices with the complete set of meson quantum numbers  were
 determined in this model  by the form of the color charged gluon propagator. The spectrum of mesons displayed the Regge character  both with respect to total angular momentum  and radial quantum number of the meson.  The  reason for confinement of a single quark and Regge spectrum of mesons turned out to be the same -- the analytic properties of quark and gluon propagators.  

This  result has almost completed the quark confinement picture based on the random almost everywhere homogeneous Abelian (anti-)self-dual fields. Self-duality of the fields plays the crucial role in this picture. This random field ensemble represents a medium where the color charged elementary excitations exist as quickly decaying in space and time field fluctuations but the collective colorless excitations (mesons) can propagate as plain waves (particles). 
It should be stressed that in this formalism any meson looks much more like  a complicated collective excitation of a medium (QCD vacuum) involving quark, antiquark and gluon fields than a nonrelativistic quantum mechanical bound state of charged  particles (quark and anti-quark) due to some potential interaction between them. Within this relativistic quantum field description the Regge spectrum of color neutral collective modes appeared as a "medium effect" as well as the suppression (confinement) of a color charged elementary modes.

However, besides this dynamical color charge confinement,  a correct complete picture must include the limit of  static quark-antiquark pair with the area law for the temporal Wilson loop. In order to explore this aspect an  explicit construction of the random domain ensemble was suggested in paper~\cite{NK1}, and the area law for the Wilson loop was demonstrated by the explicit calculation. Randomness of the ensemble (in line with \cite{Olesen7}) and (anti-)self-duality of the fields are crucial for this result.

In this paper we briefly review the approach to confinement, chiral symmetry realization and bosonization based  on the   representation of  QCD vacuum  in terms of the statistical ensemble of  almost everywhere homogeneous Abelian (anti-)self-dual gluon fields,  systematically calculate the spectrum of radial  meson excitations and their decay constants and outline  the  possible relation between the  formalism of soft wall AdS/QCD and light-front holography, and this, at first sight,   different approach.

 The  character of meson wave functions   in hadronization approach~\cite{EN1} is fixed by the form of the gluon propagator in the  background  of the  specific class of vacuum gluon configurations.     These wave functions  are almost identical to the wave functions of the soft wall AdS/QCD with  quadratic dilaton profile and Leutwyler-Stern formalism. In all three cases we are dealing with the generalized Laguerre polynomials as the functions of $z$. In the hadronization approach of \cite{EN1} and \cite{EN}  the fifth coordinate $z$ appears as the relative distance between quark and antiquark in the center of quark mass coordinate system, while the  center of mass coordinate $x$ represents the space-time point where the meson field is localized. This treatment of coordinates goes in line with the Leutwyler-Stern approach. Comparison of the soft wall AdS/QCD action and the effective action for  auxiliary bilocal meson-like fields of the hadronization approach hints at the link  between   very appearance of dilaton and its particular profile and the form of nonperturbative gluon propagator.  The strictly  quadratic in $z$ dilaton profile corresponds to the propagator in the presence of strictly homogeneous (anti-)self-dual Abelian gluon field that is an idealization of the domain wall network background with infinitely thin domain walls. These three approaches can be considered as complementary to each other ways to describe  confinement in terms of meson wave functions. However, unlike two other approaches bosonization  in the background of domain wall networks relates the  form of meson wave functions to  the particular  vacuum structure of QCD and provides one with the manifestly defined meson effective action  that  describes  strong, electromagnetic and weak interactions of mesons in terms of nonlocal vertices given by the quark-gluon loops. 
 New results for mass spectrum and decay constants of radially excited  light, heavy-light mesons and heavy quarkonia are presented. An overall accuracy of description is 10-15 percent in the lowest order calculation
 achieved with the minimal for QCD  set of parameters: infrared limits of renormalized strong coupling constant $g$ and quark masses $m_f$, scalar gluon condensate $\langle g^2F^2\rangle$ as a fundamental scale of QCD
 and topological susceptibility of pure QCD without quarks.  This last parameter can be related to the mean size of  domains.  It should be noted that the present paper also completes and clarifies the studies of \cite{EN1,EN,NK4} in two important respects: diagonalization of the quadratic part of the meson effective action with respect to radial quantum number, 
clarification of the physical meaning of the quark mass parameters  in the context of the spontaneous chiral symmetry breaking by the background field and four-fermion interaction. 
 
The paper is organized as follows. Section \ref{section_domain_wall_network_as_vacuum} is devoted to motivation of the approach.  Derivation of the effective meson action  is considered in section \ref{section_domain_model}.  Results for the  masses, transition and decay constants of various mesons are presented in  section \ref{section_masses_of_mesons}. In the  section \ref{discussion}  we outline possible relation between the present hadronization approach and the formalism of the soft wall AdS/QCD model, light front holographic QCD, compare the quark and gluon propagators of the present approach with the results of functional renormalization group (FRG) and Dyson-Schwinger equations (DSE).  Important technical details are given  in the appendices.

\section{Domain wall networks as QCD vacuum \label{section_domain_wall_network_as_vacuum}}

The primary phenomenological basis of the present approach is the existence of nonzero condensates in QCD, first of all -- the scalar gluon  condensate $\langle g^2F^2\rangle$.  In order to incorporate this condensate into the  functional integral approach to quantization of QCD one has to choose appropriate  conditions for  the functional space of gluon fields $A_\mu^a$ to be integrated over (see, e.g., Ref.\cite{faddeev}). 
Besides the formal mathematical  content, these conditions play the role of substantial physical input  which, together with the classical action of QCD, complements the statement of the quantization problem. In other words, starting with  the very basic  representation of the Euclidean  functional integral for QCD, 
\begin{equation}
\label{functional_integral}
Z=N\int\limits_{{\cal F}_B} DA \int\limits_{\Psi} D\psi D\bar\psi\exp\{-S[A,\psi,\bar{\psi}]\},
\end{equation}
one has to specify integration spaces $\mathcal{F}_B$ for gluon and  $\Psi$ for quark fields. 
Bearing in mind a nontrivial QCD vacuum structure encoded in various condensates, one have to define  $\mathcal{F}_B$  permitting  gluon fields with nonzero classical action density,
\begin{equation*}
{\cal F}_B=\left\{A: \lim_{V\to \infty}\frac{1}{V}\int_V d^4xg^2F^a_{\mu\nu} (x)F^a_{\mu\nu}(x) =B^2\right\}.
\end{equation*}
It is assumed that the   constant $B$  may have a nonzero value.  
The gauge fields $A$ that satisfy this condition  have a potential to provide the vacuum with the whole variety of  condensates.

An analytical approach to definition and calculation of the functional integral can be based on separation of modes  $B_\mu^a$ responsible for nonzero  condensates  from the small perturbations $Q_\mu^a$. This separation must be supplemented with gauge fixing.  Background gauge fixing condition $D(B)Q=0$ is the most natural choice. To perform separation, one inserts identity
\begin{equation*}
1=\int\limits_{{\cal B}}DB \Phi[A,B]\int\limits_{{\cal Q}} DQ\int\limits_{\Omega}D\omega \delta[A^\omega-Q^\omega-B^\omega]
 \delta[D(B^\omega)Q^\omega]
\end{equation*}
in the functional integral and arrives at 
\begin{eqnarray*}
Z &=&N'\int\limits_{{\cal B}}DB \int\limits_{\Psi} D\psi D\bar\psi\int\limits_{{\cal Q}} DQ \det[\mathcal{D}(B)\mathcal{D}(B+Q)]
\delta[\mathcal{D}(B)Q]e^{-S_{\rm QCD}[B+Q,\psi, \bar\psi]}\\
&=&\int\limits_\mathcal{B}DB \exp\{-S_\mathrm{eff}[B]\}.
\end{eqnarray*}
Thus defined quantum effective action $S_\text{eff}[B]$ has a physical meaning of the free energy of the quantum field system in the presence of the background  gluon field $B_\mu ^a$.  In the limit $V\to \infty$  global minima of  $S_\text{eff}[B]$ determine the class of  gauge field configurations  representing  the equilibrium state (vacuum) of the system. 

Quite reliable argumentation in favour of (almost everywhere) homogeneous Abelian  (anti-)self-dual fields as dominating vacuum configurations was put forward by many authors~\cite{Pagels,Leutwyler2}.  
As it has already been mentioned in Introduction, nonperturbative calculation of QCD quantum effective action  within the functional renormalization group approach \cite{Pawlowski} supported the  one-loop result \cite{Pagels,Minkowski,Leutwyler2} and indicated the existence of a minimum of the effective potential for nonzero value of Abelian (anti-)self-dual homogeneous gluon field.

\begin{figure}
\begin{centering}
\includegraphics[width=0.35\textwidth]{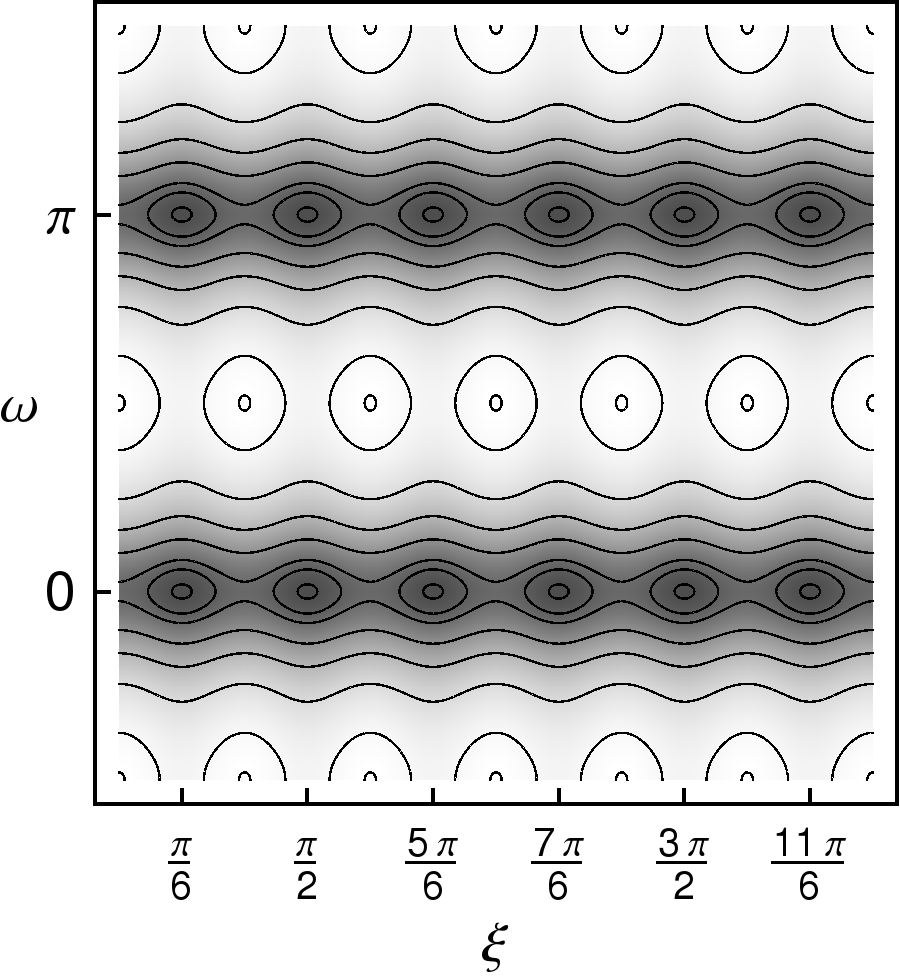}
\end{centering}
\caption{Effective potential \eqref{ueff} as a function of the angle $\omega$ between chromomagnetic and chromoelectric field  and the mixing angle $\xi$  in the Cartan subalgebra. The minima in the dark gray regions correspond to the Abelian  (anti-)self-dual  configurations and form a periodic  structure labelled  by integer indices $(kl)$  in Eq.~\eqref{minima} (for more details see  \cite{NK1,NG2011,NV}).\label{effpotxiomega}}
\end{figure}

Ginzburg-Landau (GL) approach to the quantum effective action indicated a possibility of the domain wall network formation in  QCD vacuum resulting in the  dominating vacuum gluon configuration seen as an ensemble of densely  packed lumps of covariantly constant Abelian (anti-)self-dual field \cite{NK1,NG2011,NV,George:2012sb}. Nonzero scalar gluon condensate $\langle g^2F^a_{\mu\nu}F^a_{\mu\nu}\rangle$ 
postulated by the effective potential
\begin{eqnarray}
U_{\mathrm{eff}}&=&\frac{\Lambda^4}{12} {\rm Tr}\left(C_1\breve{ f}^2 + \frac{4}{3}C_2\breve{ f}^4 - \frac{16}{9}C_3\breve{ f}^6\right),
\label{ueff}
\end{eqnarray}
with  $\Lambda$ being a scale of QCD  and $\breve f_{\mu\nu}=\breve T^aF^a_{\mu\nu}/\Lambda^2$,
leads 
to the existence  of twelve discrete degenerate global minima of the effective action (see Fig.\ref{effpotxiomega}), 
\begin{eqnarray}
&&\breve A_{\mu}\in\left\{\breve B^{(kl)}_{\mu}| \  k=0,1,\dots,5; \  l=0,1\right\}, \ \
\breve B^{(kl)}_{\mu}  = -\frac{1}{2}\breve n_k B^{(l)}_{\mu\nu}x_\nu, 
\nonumber\\
 && \tilde B^{(l)}_{\mu\nu}=\frac{1}{2}\varepsilon_{\mu\nu\alpha\beta} B^{(l)}_{\alpha\beta}=(-1)^l B^{(l)}_{\mu\nu},
\nonumber\\
&&\breve n_k = T^3\ \cos\left(\xi_k\right) + T^8\ \sin\left(\xi_k\right),
\ \
\xi_k=\frac{2k+1}{6}\pi,
\label{minima}
\end{eqnarray}
where $l=0$ and $l=1$ correspond to the self-dual and anti-self-dual field respectively,  matrix $\breve{n}_k$ belongs to Cartan subalgebra of $su(3)$ with six values of the angle $\xi_k$ corresponding to the boundaries of the Weyl chambers in the root space of $su(3)$. 

The minima are connected by the parity and Weyl group reflections.  
Their existence indicates that the system is prone to the domain wall formation.  To demonstrate the simplest example of domain wall interpolating between the self-dual and anti-self-dual Abelian configurations, one allows  the angle  $\omega$ between chromomagnetic and chromoelectric fields to vary from point to point in $R^4$  and restricts other degrees of freedom of gluon field to their vacuum values.
In this case Ginsburg-Landau Lagrangian  leads  to the  sine-Gordon equation for  $\omega$ with the standard
 kink solution (for details see Ref.~\cite{NG2011,NV})
\begin{equation*}
 \omega(x_\nu) = 2\ \arctan \left(\exp(\mu x_\nu)\right).
\end{equation*}

\begin{figure}[h!]
\begin{center}
\includegraphics[width=.6\textwidth]{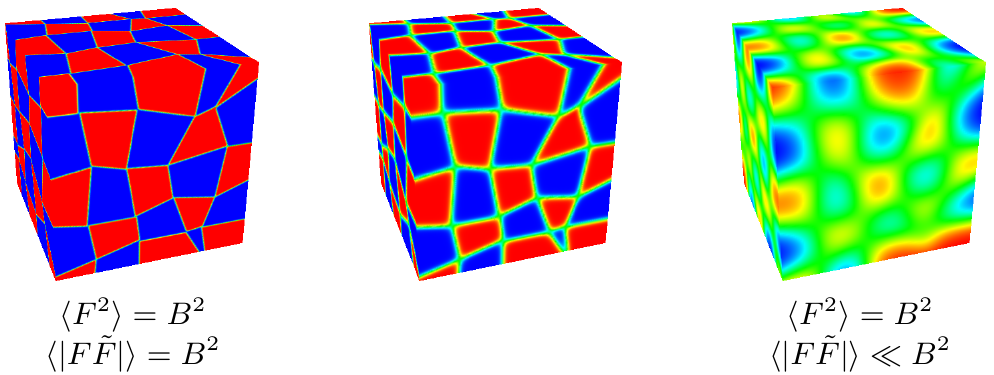}
\end{center}
\caption{Topological charge density for domain wall networks with different values of the wall width $\mu$. The leftmost picture is an example of confining almost everywhere homogeneous Abelian (anti-)self-dual fields. Red (blue) color corresponds to the self-dual field (anti-self-dual), green -- pure chromomagnetic field. The rightmost plot represents the case of preferably pure chromomagnetic field when the topological charge density is nearly zero and color charged quasiparticles can be excited thus indicating deconfinement (for more details see \cite{NV}).\label{cubes}}
\end{figure}

Away from the kink location vacuum  field is almost self-dual ($\omega=0$) or anti-self-dual ($\omega=\pi$).  Exactly at the wall it becomes purely chromomagnetic ($\omega=\pi/2$).  Domain wall network is constructed  by means of the kink superposition.
In general kink can be parametrized as 
\begin{equation*}
 \zeta(\mu_i,\eta_\nu^{i}x_\nu-q^{i})=\frac{2}{\pi}\arctan\exp(\mu_i(\eta_\nu^{i}x_\nu-q^{i})),
\end{equation*}
where $\mu^i$ is inverse width of the kink, $\eta_\nu^i$ is a normal to the wall and $q^i=\eta_\nu^i x_\nu$ are coordinates of the wall.
A single lump in two, three and four dimensions is given by
\begin{equation*}
\omega(x)=\pi\prod_{i=1}^k \zeta(\mu_i,\eta_\nu^{i}x_\nu-q^{i}).
\end{equation*}
for $k=4,6,8$, respectively.
The general kink network is then given by the additive superposition of lumps
\begin{equation*}
\omega=\pi\sum_{j=1}^{\infty}\prod_{i=1}^k \zeta(\mu_{ij},\eta_\nu^{ij}x_\nu-q^{ij}).
\end{equation*}
Topological charge density distribution for a network of domain walls with different width is illustrated in Fig.\ref{cubes}. 

Based on this construction, the measure of integration over the background field $B_\mu^a$ can be constructively represented as the infinite dimensional (in the infinite volume) integral over the parameters  of $N\to\infty$ domain walls in the network:  their positions, orientations and widths, with the weight  determined by the effective action.  It should be noted that chronologically the  explicit construction of the  domain wall network is the most recent development of the 
formalism that have been studied in the series of papers \cite{EN,EN1,NK1,NK4,NK6}, in which the domain wall defects in the homogeneous Abelian (anti-)self-dual field were taken into account either implicitly or 
in an  explicit but simplified form with the spherical domains.  The practical calculations in the next sections will be done within combined implementation of domain model given in paper~\cite{NK4}: propagators in the quark loops are taken in the approximation of the homogeneous background field and the quark loops are averaged over    the background field,  the correlators of the  background field  are calculated in the spherical domain approximation.

\section{Hadronization within the domain model of QCD vacuum \label{section_domain_model}}

The haronization formalism based on domain model of QCD vacuum  was elaborated in the series of papers  \cite{EN1,EN,NK1,NK4}. 
We refer to these papers for most of the technical details omitted in this brief presentation.  It has been shown that the  model  embraces static (area law) and dynamical quark confinement (propagators in momentum representation  are entire analytical functions) as well as spontaneous breaking of chiral symmetry by the background domain structured field itself. $U_A(1)$ problem was resolved without introducing the strong CP-violation~\cite{NK6}. Estimation of masses of light, heavy-light mesons and heavy quarkonia along with their orbital excitations \cite{EN1,EN,NK4} demonstrated	promising phenomenological performance.  However, calculations in Refs.~\cite{EN,NK4} have been done neglecting  a mixing between  radially excited meson fields.  Below we present results of  calculation  refined in this respect.

In the spherical domain approximation, the background gluon  fields are represented by the ensemble of domain-structured fields with the strength tensor \cite{NK1,NK4}
\begin{gather}
\nonumber
F^{a}_{\mu\nu}(x)
=\sum_{k=1}^N n^{(k)a}B^{(k)}_{\mu\nu}\theta(1-(x-z_k)^2/R^2), \quad B^{(k)}_{\mu\nu}B^{(k)}_{\mu\rho}=B^2\delta_{\nu\rho}, \quad B=\frac{2}{\sqrt{3}}\Lambda^2,
\\\nonumber
\tilde B^{(k)}_{\mu\nu}=\pm B^{(k)}_{\mu\nu}, \quad 
\hat n^{(k)}=t^3\cos\xi_k+t^8\sin\xi_k, \quad \xi_k\in\left\{\frac{\pi}{6}(2k+1),\ k=0,\dots,5\right\},
\end{gather}
where $z_k$ is the space-time coordinate of the $k$-th domain center,  scale $\Lambda$ and  mean domain radius $R$  are parameters of the model related to the
scalar gluon condensate and topological susceptibility  of pure Yang-Mills vacuum, respectively~\cite{NK1}.

The measure of  integration over ensemble of background fields is defined as~\cite{NK1,NK4}
\begin{eqnarray*}
\int\limits_{\cal B}dB\dots  &=& \prod_k
\frac{1}{24\pi^2}\lim_{V\to\infty}\frac{1}{V}
\int_V d^4z_k
\int\limits_0^{2\pi}d\varphi_k\int_0^\pi d\theta_k\sin\theta_k
\\
&\times&\int_0^{2\pi} d\xi_k\sum\limits_{l=0,1,2}^{3,4,5}
\delta\left(\xi_k-\frac{(2l+1)\pi}{6}\right)
\int_0^\pi d\omega_k\sum\limits_{n=0,1}\delta(\omega_k-\pi n)
\dots 
\end{eqnarray*}

Once the measure  is specified, one can return to the functional integral \eqref{functional_integral} and integrate out fluctuation part of the gluon fields $Q$:
\begin{eqnarray*}
{\cal Z} & =&  
\int dB \int_{\Psi}{\cal D}\psi {\cal D}\bar \psi
\int_{{\cal Q}} {\cal D}Q \delta[D( B)Q]
\Delta_{\rm FP}[B,Q]
e^{
- S^{\rm QCD}\left[Q+ B,\psi,\bar\psi
\right]} \\
&=&
\int dB \int_{\Psi}{\cal D}\psi {\cal D}\bar \psi
\exp\left\{\int dx \bar\psi\left(i \hspace*{-0.2em} \not{\hspace*{-0.2em}\partial} + g\hspace*{-0.3em}\not{\hspace*{-0.3em}B}-m\right)\psi\right\} W[j],
\end{eqnarray*}
where $j^a_\mu(x) =\bar\psi(x) \gamma_\mu t^a \psi(x)$ is the local quark current.  Recalling the definition of Green functions,
\begin{equation*}
G^{a_1\dots a_n}_{\mu_1\dots \mu_n}(x_1,\dots,x_n|B)=\left. \frac{1}{g^n}\frac{\delta^n \ln W[j]}{\delta j^{a_1}_{\mu_1}(x_1)\dots \delta j^{a_n}_{\mu_n}(x_n)} \right\lvert_{j=0},
\end{equation*}
we arrive at the representation 
\begin{eqnarray*}
W[j|B]&=&\exp\left\{\sum_n \frac{g^n}{n!}\int d^4x_1\dots \int d^4x_n j^{a_1}_{\mu_1}(x_1)\dots j^{a_n}_{\mu_n}(x_n) G^{a_1\dots a_n}_{\mu_1\dots \mu_n}(x_1,\dots,x_n|B)\right\},
\end{eqnarray*}
where by construction the gauge coupling constant $g$  and the  exact renormalized  $n$-point gluon Green functions of pure gauge theory in the presence of the background field $B$  appear  to be renormalized within appropriate renormalization scheme.  It is needless to say that the functional form of these Green functions,   gluon propagator in particular, has been a subject of many investigations carried out over decades.  Quite reliable information about two- and three-point Euclidean Green functions was obtained within the functional renormalization group, Lattice QCD as well as calculations based on Dyson-Schwinger equations.  

At this step one has to set up the  approximation scheme. We truncate the exponent in $W[j|B]$ up to the four-fermion interaction term. Interaction between standard local color charged quark currents is described by  the product of the renormalized coupling constant squared and  exact gluon propagator  $g^2G^{a_1 a_2}_{\mu_1 \mu_2}(x_1,x_2|B)$ which will be approximated  by the gluon propagator in the presence of homogeneous Abelian (anti-)self-dual field.  Radiative corrections due to the gluon and ghost field fluctuations are neglected  (for more details see Refs.~\cite{EN1,EN,NK4}).  It should be noted that omitted radiative corrections can be represented in terms of the standard for pure gluodynamics set of Feynman graphs for gluon polarization function but the internal lines in the graphs correspond to the gluon and ghost propagators in the background  field $B$.  In other words,  the  approximation  in use corresponds to  the lowest  (tree level) order  with respect to perturbative fluctuations $Q$, but the  background field (vacuum field $B$) itself  is taken into account exactly. 

The randomness  of domain ensemble is taken into account implicitly by means of  averaging the nonlocal meson-meson interaction vertices over all possible configurations of the homogeneous background field at the final stage of derivation of the effective meson action~\cite{EN,NK4}.

Relevant truncated part of QCD functional integral reads
\begin{eqnarray}
\mathcal{Z}=\int_{\cal B} dB \int_{\Psi}{\cal D}\psi {\cal D}\bar \psi
\exp\left\{\int d^4x \bar\psi\left(i \hspace*{-0.2em} \not{\hspace*{-0.2em}\partial} + g\hspace*{-0.3em}\not{\hspace*{-0.3em}B}-m\right)\psi+\mathcal{L}\right\},
\label{4ferm}\\
\mathcal{L}=\frac{g^2}{2}\int d^4x \int d^4y\  G^{a b}_{\mu \nu}(x,y|B)j^{a}_{\mu}(x) j^{b}_{\nu}(y),
\nonumber
\end{eqnarray}
where $m$ is a diagonal quark mass matrix.  By means of the standard Fierz transformation of color, Dirac and flavour matrices the four-quark interaction can be rewritten as
 \begin{eqnarray}
\mathcal{L}=\frac{g^2}{2}\sum_{J,c}C_J\int d^4x \int d^4y  G(x-y)J^{Jc}(x,y|B)J^{Jc}(y,x|B) ,
\nonumber
\end{eqnarray}
 where numerical coefficients $C_J$ are different for different spin-parity  $J=S,P,V,A$. Here bilocal color neutral quark currents,
 \begin{equation*}
J^{Jc}(x,y|B)= \bar\psi (x)\lambda^c\Gamma_J\exp\left\{\frac{ i}{2}x_\mu \hat B_{\mu\nu}y_\nu \right\}\psi(y),
\end{equation*}
are singlets with respect to the local background gauge transformations.   In the center of  quark mass coordinate system bilocal currents take the form
 \begin{eqnarray}
&&J^{Jc}(x,y|B)\to J^{Jc}(x,z|B)=\bar\psi_f(x)\lambda^c\Gamma_J
\exp\left( iz_\mu  \stackrel{\hspace{-4.5mm}\leftrightarrow}{\mathcal{D}_{ff'}^\mu}(x) \right)\psi_{f'}(x),
\label{biloccur}\\
&&\stackrel{\hspace{-4.5mm}\leftrightarrow}{\mathcal{D}^{ff'}_{\mu}}=\xi_f\stackrel{\leftarrow}{\mathcal{D}}_{\mu}-\xi_{f'}\stackrel{\rightarrow}{\mathcal{D}}_{\mu}, 
\ \
\stackrel{\leftarrow}{\mathcal{D}}_{\mu}(x)=\stackrel{\leftarrow}{\partial}_\mu+i\hat B_\mu(x),  \ \ 
\stackrel{\rightarrow}{\mathcal{D}}_{\mu}(x)=\stackrel{\rightarrow}{\partial}_\mu-i\hat B_\mu(x), 
\nonumber\\
&& \xi_f=\frac{m_{f'}}{m_f+m_{f'}},\ \xi_{f'}=\frac{m_{f}}{m_f+m_{f'}},
\nonumber
\end{eqnarray}
and  their interaction is described by the action~\cite{EN1}
 \begin{eqnarray}
&&\mathcal{S}=\frac{g^2}{2}\sum_{J,c}C_J\int d^4x \int d^4z G(z)J_{Jc}^{\dagger}(x,z|B)J_{Jc}(x,z|B),
\label{4ferm-1}\\
&& G(z)=\frac{1}{4\pi^2 z^2}\exp\left\{-\frac{1}{4}\Lambda^2z^2\right\}, 
\label{glprop-1}
\end{eqnarray}

\noindent
where $x_\mu$ - center of quark mass coordinates, and $z_\mu$ - relative coordinates of quark and antiquark.
It has to be noted here that  quark fields are seen as pure fluctuations   describable in terms of four-dimentional harmonic oscillator eigenmodes of the bound state type~\cite{NV,NK1,NK4} in $R^4$. Interpretation of the quark field in terms of point-like particle is simply does not exist in the confining background under consideration.
Function   $G(z)$ originates from the gluon propagator in the presence of the homogeneous Abelian (anti-)self-dual gluon field~\cite{EN1}. It differs from the free massless scalar propagator by the Gaussian exponent,
which completely changes the IR properties of the propagator but leaves its UV asymptotic behaviour  unchanged. In momentum representation it takes the form
\begin{eqnarray}
\label{gprop}
\tilde G\left(p\right) = \frac{1}{p^2}\left(1-e^{-p^2/\Lambda^2}\right).
\end{eqnarray}
It is important that nonzero gluon condensates $\langle g^2F^2\rangle$ and $\langle g^2|\tilde FF|\rangle$ represented by the Abelian (anti-)self-dual  vacuum field  remove the pole from the propagator which can be treated as dynamical confinement of the color charged fields~\cite{Leutwyler1}.

\begin{figure}[h!]
\begin{center}
\includegraphics[scale=1]{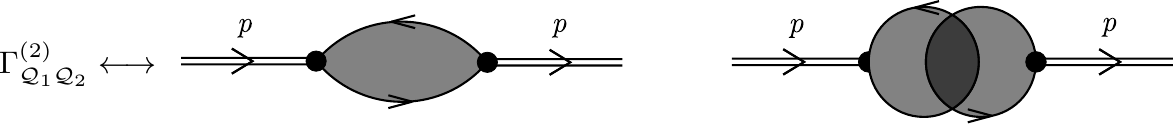}\\
\includegraphics[scale=1]{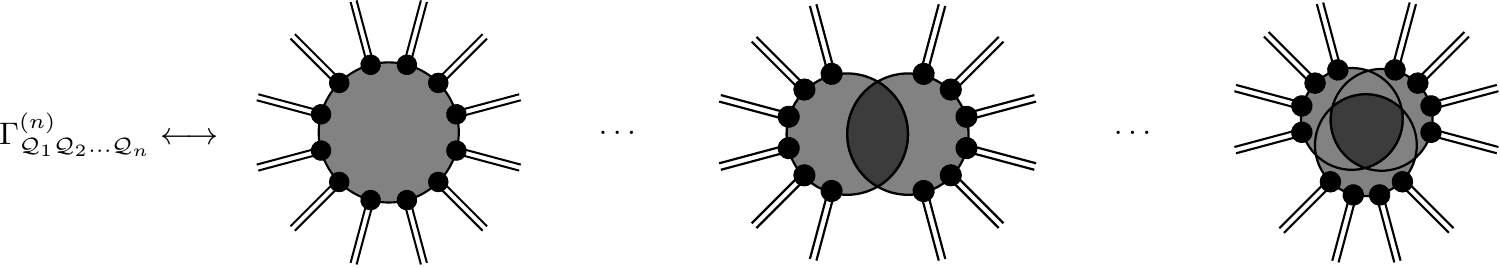}\end{center}
\caption{Diagrammatic representation of nonlocal meson vertex functions. Light grey denotes averaging over background field, dark grey denotes correlation of loop diagrams by background field.\label{diagrams}}
\end{figure}

 The  quark propagator in the homogeneous as well as domain structured~\cite{NK1} Abelian (anti-)self-dual gluon field also demonstrates confinement.
Momentum representation $\tilde H_f(p|B)$ of the translation invariant part of the quark propagator in the presence of the homogeneous  field, 
\begin{equation*}
S(x,y)=\exp\left(-\frac{i}{2}x_\mu B_{\mu\nu}y_\nu\right)H(x-y),
\end{equation*}
is an entire analytical function of momentum:
\begin{eqnarray}
\label{qprop}
\tilde H_f(p)=\frac{1}{2 v  \Lambda^2} \int_0^1 ds e^{(-p^2/2 v  \Lambda^2)s}\left(\frac{1-s}{1+s}\right)^{m_f^2/4 v  \Lambda^2}\left[\vphantom{\frac{s}{1-s^2}}p_\alpha\gamma_\alpha\pm is\gamma_5\gamma_\alpha f_{\alpha\beta} p_\beta \right.
\nonumber\\
\left.\mbox{}+m_f\left(P_\pm+P_\mp\frac{1+s^2}{1-s^2}-\frac{i}{2}\gamma_\alpha f_{\alpha\beta}\gamma_\beta\frac{s}{1-s^2}\right)\right],\\
f_{\alpha\beta}=\frac{\hat{n}}{2v\Lambda^2}B_{\alpha\beta}, \   v =\mathrm{diag}\left(\frac16,\frac16,\frac13\right), \ \ \hat{B}_{\rho\mu}\hat{B}_{\rho\nu}=4 v^2\Lambda^4\delta_{\mu\nu}.
\nonumber
\end{eqnarray}

The propagator has a rich  Dirac structure including not only the vector and scalar parts but also the pseudoscalar, axial vector and tensor terms (flavour index $f$ is omitted for the sake of brevity)
\begin{eqnarray}
\label{qpropSPVAT}
\tilde H(p)=\frac{m}{2 v \Lambda^2} {\cal H}_S(p^2) \mp\gamma_5\frac{m}{2 v \Lambda^2} {\cal H}_P(p^2)+
\gamma_\alpha\frac{p_\alpha}{2 v \Lambda^2} {\cal H}_V(p^2) \pm i\gamma_5\gamma_\alpha\frac{ f_{\alpha\beta}p_\beta}{2 v \Lambda^2} {\cal H}_A(p^2)+\sigma_{\alpha\beta}\frac{m f_{\alpha\beta} }{4 v \Lambda^2} {\cal H}_T(p^2).
\end{eqnarray}
Here "$\pm$" corresponds to self-dual and anti-self-dual background field configurations. One can easily reconstruct explicit form of  functions $\mathcal{H}_J$ from Eq.\eqref{qprop}.   
More detailed description of different form factors, particularly the scalar one, and their role in the chiral symmetry realization will be given in  section~\ref{discussion}.
This structure of the quark propagator plays important role for successful description of the meson spectrum,
especially for the ground state light mesons.  

\begin{table}[htbp]
\caption{Model parameters fitted to the masses of $\pi,\rho,K,K^*, \eta', J/\psi,\Upsilon$ and used in  calculation of all other meson masses, decay and transition constants, for $N=7$ (see explanations in the text and Fig.\ref{deltaPN}).}
{\begin{tabular}{@{}ccccccc@{}}
 \toprule
$m_{u/d}$, MeV&$m_s$, MeV&$m_c$, MeV&$m_b$, MeV&$\Lambda$, MeV&$\alpha_s$&$R$, fm\\
\colrule
$145$&$376$&$1566$&$4879$&$416$&$3.45$&$1.12$\\
\botrule
\end{tabular}}
\label{values_of_parameters}
\end{table}

There are two equivalent ways to derive effective meson action based on the functional integral  \eqref{4ferm} with the interaction term $\mathcal{L}$  taken in the form~\eqref{4ferm-1}.  The first one is bosonization of the functional integral in terms of bilocal meson-like fields (see for example Ref.~\cite{roberts}). We shall return to this option in the discussion section. Another, more elucidative  way is to decompose the bilocal currents \eqref{biloccur}
over complete set of  functions $f^{nl}_{\mu_1\dots \mu_l}(z)$ orthogonal with  the weight determined by function $G(z)$ originating from the gluon propagator  \eqref{glprop-1} in Eq.\eqref{4ferm-1}
\begin{equation*}
J^{aJ}(x,z)=\sum_{n,l=0}^{\infty}\left(z^2\right)^{l/2} f^{nl}_{\mu_1\dots \mu_l}(z) J^{aJln}_{\mu_1\dots \mu_l}(x).
\end{equation*}
Here  $n$ is the radial quantum number and  $l$  is the orbital momentum.
Coefficient quark currents  $J^{aJln}_{\mu_1\dots \mu_l}(x)$  have to describe  intrinsic structure of the collective meson-like excitations with complete set of  quantum numbers.  The form of interaction~\eqref{4ferm-1} and natural requirement of  diagonality (with respect to  $n$ and $l$) of the four-quark interaction,  expressed in terms of the currents $J^{aJln}_{\mu_1\dots \mu_l}(x)$,  indicate the
 choice of  $f^{nl}(z)$ 
\begin{equation}
f^{nl}_{\mu_1\dots \mu_l}=L_{nl}\left(z^2\right) T^{(l)}_{\mu_1\dots \mu_l}(n_z),\quad n_z=z/\sqrt{z^2}.
\label{wavef}
\end{equation}
Here $T^{(l)}_{\mu_1\dots \mu_l}$ are irreducible tensors of four-dimensional rotational group, and 
generalized Laguerre polynomials $L_{nl}$  obey relation
\begin{equation*}
\int_0^\infty du \rho_l(u)L_{nl}(u)L_{n'l}(u)=\delta_{nn'},\quad \rho_l(u)=u^{l}e^{-u}.
\end{equation*}

The weight $\rho_l(u)$ comes from the gluon propagator~\eqref{glprop-1}.  Nonlocal quark currents $J^{aJln}_{\mu_1\dots \mu_l}$ with  complete set of meson quantum numbers  can be explicitly calculated and depend only on the center of mass coordinate $x$~\cite{EN,NK4},
\begin{eqnarray}
J^{aJln}_{\mu_1\dots \mu_l}(x)=\bar{q}(x)V^{aJln}_{\mu_1\dots \mu_l}\left(\frac{\stackrel{\leftrightarrow}{D} \!
(x)}{\Lambda}\right)q(x),
\nonumber\\
V^{aJln}_{\mu_1\dots\mu_l}(x)= {\cal C}_{Jln}M^a\Gamma^J F_{nl}\left(\frac{\stackrel{\leftrightarrow}{\cal D}^2\!\!\!
(x)}{\Lambda^2}\right)T^{(l)}_{\mu_1\dots\mu_l}\left(\frac{1}{i}\frac{\stackrel{\leftrightarrow}{\cal D}\!(x)}{\Lambda}\right),
\label{qmvert}\\
F_{nl}(s)=s^n\int_0^1 dt t^{n+l} \exp(st),
\nonumber\\
 {\cal C}^2_{Jln}=C_J\frac{l+1}{2^ln!(n+l)!},\quad  C_{S/P}=2C_{V/A}=\frac{1}{9},
\nonumber
\end{eqnarray}
where $M^a$ and $\Gamma^J$ are flavour $SU(N_f)$ and Dirac matrices respectively.
The four-fermion interaction takes the form of an infinite sum of the current-current interactions diagonal with respect to all quantum numbers
\begin{equation*}
\mathcal{L}=\frac{g^2}{2} \sum_{aJln} \int d^4x J_{aJln}^{\dagger}(x)J_{aJln}(x).
\end{equation*}
It has to be stressed that the nonlocal quark currents are invariant with respect to the local gauge transformations of the background gauge field as the vertices \eqref{qmvert} depend on the covariant derivatives.

\begin{figure}
\begin{centering}
\includegraphics[width=.35 \textwidth]{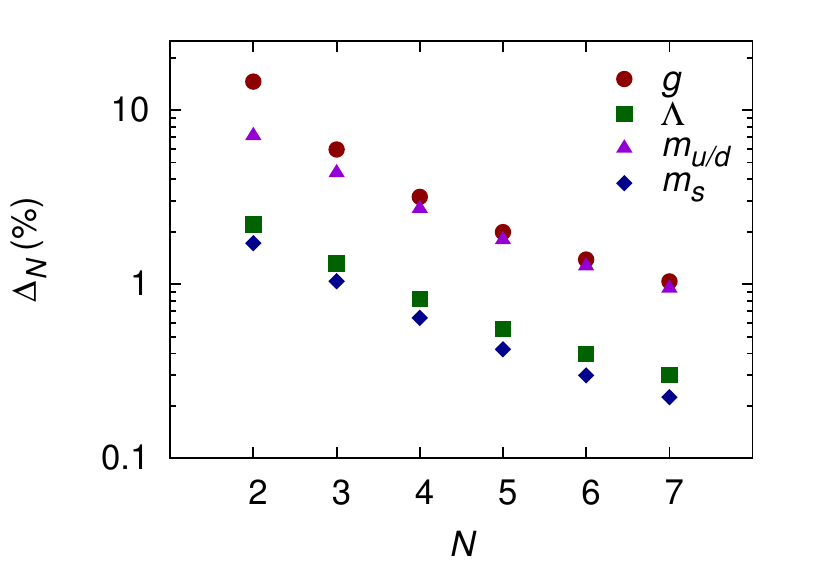}  
\end{centering}
\caption{ Dependence of relative variation of the model parameters $\Delta_N=|P_{N}-P_{N-1}|/P_{N-1}$ on the number of Laguerre polynomials $N$ taken into account during  diagonalization of the quadratic part of the meson action with respect to the radial number $n$.  
It can be  seen that the iterations converge with $N$ for all quantities faster than $\Delta_N\approx \Delta_1\exp(-.35N)$ for $N>5$.
For $N=7$  the variations of light quark mass $m_{u/d}$ and coupling constant $g$ slow down to one percent level,  while the change of the scale $\Lambda$ and the strange quark mass $m_s$ approach a  fraction of a percent.   
Alternations of the masses of heavy $c$ and $b$  quarks  and domain size $R$ are not shown as it is of order of $.1\%$ or less for any $N$ which is far 	smaller than the overall accuracy of the model.\label{deltaPN}}
\end{figure}

 The truncated QCD functional integral can be rewritten in terms of the   composite colorless meson fields $\phi_{\cal Q}$ by means of  the standard bosonization procedure: introduce the auxiliary meson fields, integrate out the quark fields, perform the orthogonal transformation of the auxiliary fields that diagonalizes the quadratic part of the action and, finally, rescale the  meson fields to provide the correct residue of the meson propagator at the pole corresponding to its physical mass (if any).  
More details can be found in Ref.~\cite{EN,EN1,NK4}. 
The result can be written in the following compact form
\begin{eqnarray}
&&Z={\cal N}
\int D\phi_{\cal Q}
\exp\left\{-\frac{\Lambda^2}{2}\frac{h^2_{\cal Q}}{g^2}\int d^4x 
\phi^2_{\cal Q}(x)
-\sum\limits_{k=2}^\infty\frac{1}{k}W_k[\phi]\right\},
\label{meson_pf}\\
&&W_k[\phi]=
\sum\limits_{{\cal Q}_1\dots{\cal Q}_k}h_{{\cal Q}_1}\dots h_{{\cal Q}_k}
\int d^4x_1\dots\int d^4x_k
\Phi_{{\cal Q}_1}(x_1)\dots \Phi_{{\cal Q}_k}(x_k)
\Gamma^{(k)}_{{\cal Q}_1\dots{\cal Q}_k}(x_1,\dots,x_k),
\label{effective_meson_action}
\\
 &&\Phi_{{\cal Q}}(x)=\int \frac{d^4p}{(2\pi)^4}e^{ipx}{\mathcal O}_{{\mathcal Q}{\mathcal Q}'}(p)\tilde\phi_{{\mathcal Q}'}(p),
 \label{Pphi}
\end{eqnarray}
where condensed index $\mathcal{Q}$ denotes all relevant meson quantum numbers and indices. Integration variables $\phi_{\mathcal Q}$ in the functional integral \eqref{meson_pf} correspond to the physical meson fields that diagonalize the quadratic part of the effective meson action \eqref{effective_meson_action} in momentum representation, which is achieved by means of orthogonal transformation ${\mathcal O}(p)$.

Interactions between physical meson fields  $\phi_{\mathcal Q}$ are  described  by  $k$-point nonlocal vertices $\Gamma^{(k)}_{\mathcal{Q}_1\dots \mathcal{Q}_2}$,
\begin{gather*}
\Gamma^{(2)}_{{\cal Q}_1{\cal Q}_2}=
\overline{G^{(2)}_{{\cal Q}_1{\cal Q}_2}(x_1,x_2)}-
\Xi_2(x_1-x_2)\overline{G^{(1)}_{{\cal Q}_1}G^{(1)}_{{\cal Q}_2}},
\\\nonumber
\Gamma^{(3)}_{{\cal Q}_1{\cal Q}_2{\cal Q}_3}=
\overline{G^{(3)}_{{\cal Q}_1{\cal Q}_2{\cal Q}_3}(x_1,x_2,x_3)}-
\frac{3}{2}\Xi_2(x_1-x_3)
\overline{G^{(2)}_{{\cal Q}_1{\cal Q}_2}(x_1,x_2)
G^{(1)}_{{\cal Q}_3}(x_3)}
\\
+
\frac{1}{2}\Xi_3(x_1,x_2,x_3)
\overline{G^{(1)}_{{\cal Q}_1}(x_1)G^{(1)}_{{\cal Q}_2}(x_2)
G^{(1)}_{{\cal Q}_3}(x_3)},
\\
\Gamma^{(4)}_{{\cal Q}_1{\cal Q}_2{\cal Q}_3{\cal Q}_4}=
\overline{G^{(4)}_{{\cal Q}_1{\cal Q}_2{\cal Q}_3{\cal Q}_4}
(x_1,x_2,x_3,x_4)}-
\frac{4}{3}\Xi_2(x_1-x_2)
\overline{G^{(1)}_{{\cal Q}_1}(x_1)
G^{(3)}_{{\cal Q}_2{\cal Q}_3{\cal Q}_4}(x_2,x_3,x_4)}
\nonumber\\\nonumber
-
\frac{1}{2}\Xi_2(x_1-x_3)
\overline{G^{(2)}_{{\cal Q}_1{\cal Q}_2}(x_1,x_2)
G^{(2)}_{{\cal Q}_3{\cal Q}_4}(x_3,x_4)}
\\\nonumber
+
\Xi_3(x_1,x_2,x_3)
\overline{G^{(1)}_{{\cal Q}_1}(x_1)G^{(1)}_{{\cal Q}_2}(x_2)
G^{(2)}_{{\cal Q}_3{\cal Q}_4}(x_3,x_4)}
\\
-\frac{1}{6}
\Xi_4(x_1,x_2,x_3,x_4)
\overline{G^{(1)}_{{\cal Q}_1}(x_1)G^{(1)}_{{\cal Q}_2}(x_2)
G^{(1)}_{{\cal Q}_3}(x_3)G^{(1)}_{{\cal Q}_4}(x_4)},
\end{gather*}
subsequently tuned to the physical meson representation by means of  corresponding orthogonal transformations ${\mathcal O}(p)$. 
Vertices $\Gamma^{(k)}$ are expressed \textit{via}  1-loop  diagrams  $G^{(k)}_{{\cal Q}_1\dots{\cal Q}_k}$ which include  nonlocal quark-meson vertices \eqref{qmvert} and quark propagators \eqref{qprop} :
\begin{equation*}
\overline{G^{(k)}_{{\cal Q}_1\dots{\cal Q}_k}(x_1,\dots,x_k)}
=\int d B
{\rm Tr}V_{{\cal Q}_1}\left(x_1|B\right)S\left(x_1,x_2\right)\dots
V_{{\cal Q}_k}\left(x_k|B\right)S\left(x_k,x_1\right),
\end{equation*}
\begin{multline*}
\overline{G^{(l)}_{{\cal Q}_1\dots{\cal Q}_l}(x_1,\dots,x_l)
G^{(k)}_{{\cal Q}_{l+1}\dots{\cal Q}_k}(x_{l+1},\dots,x_k)}
=
\\
\int d B
{\rm Tr}\left\{
V_{{\cal Q}_1}\left(x_1|B\right)S\left(x_1,x_2|B\right)\dots
V_{{\cal Q}_k}\left(x_l|B\right)S\left(x_l,x_1|B\right)
\right\}\\
\times
{\rm Tr}\left\{
V_{{\cal Q}_{l+1}}\left(x_{l+1}|B\right)S\left(x_{l+1},x_{l+2}|B\right)\dots
V_{{\cal Q}_k}\left(x_k|B\right)S\left(x_k,x_{l+1}|B\right)
\right\}.
\end{multline*}
Bar denotes integration over all configurations of the background fields.   As it is illustrated in 
Fig.\ref{diagrams},  vertex functions $\Gamma^{(k)}$ include, in general,  several  one-loop diagrams  correlated via the background field.  In the simplified model of spherical domains, the $n$-point correlator $\Xi_n(x_1,\dots,x_n)$ is given by a volume of overlap of $n$ four-dimensional hyperspheres~\cite{NK1,NK4}.

It has to be noted that though all Dirac structures, besides the vector and scalar ones, are nullified by the integration of the propagator \eqref{qpropSPVAT} over the background field,  all of them give highly nontrivial contribution to the quark loops (products of several propagators). For example, the two-point correlators responsible for the mass spectrum contain not only the one gluon (fluctuation $Q$!) exchange interaction hidden in the vertex $V_{\cal Q}$
but also additional  $S$, $P$, $V$, $A$ and $T$  interactions effectively generated by the background gluon field $B$.   

It has to be stressed that the terms linear in meson fields are absent in \eqref{meson_pf}. The linear terms naturally vanish for all mesons besides the scalar ones, and their elimination   for the scalar fields requires solution of  an infinite system of equations
\begin{eqnarray}
\Lambda^2\Phi_{{\cal Q}_1}^{(0)}=\sum\limits_{k=1}^\infty\frac{g^k}{k}\sum\limits_{{\cal Q}_1\dots{\cal Q}_k}
\Phi^{(0)}_{{\cal Q}_2}\dots \Phi^{(0)}_{{\cal Q}_k}
\Gamma^{(k)}_{{\cal Q}_1\dots{\cal Q}_k},
\label{condensates}
\end{eqnarray}
where ${\cal Q}_k=\{a_kS0n_k\}$ and $\Phi^{(0)}_{{\cal Q}_k}=\mathrm{const}$ can be treated as an infinite set of  scalar quark condensates labelled by the radial quantum number $n$.
As we shall discuss in  section~\ref{discussion}, solution of this system of equations  leads to the interesting details of the chiral symmetry realization in the presence of the background field under consideration. Actual calculations further below will be done with constant mass which from now on will be treated as the infrared limit of the running nonperturbative  quark masses $m_f(0)$ considered as  parameters of the model.

\begin{table}[ph]
\caption{Masses of light mesons. $\tilde M$ denotes the value in the chiral limit.}
{\begin{tabular}{@{}cccccc|cccccc@{}} \toprule
Meson  & $n$ & $M_{\rm exp}$\cite{PDG} & $M$ & \rule{0ex}{1.2em}$\tilde{M}$ &$h$&
 Meson & $n$ & $M_{\rm exp}$\cite{PDG} & $M$ & \rule{0ex}{1.2em}$\tilde{M}$ &$h$\\
       &     &( MeV)                   & (MeV)& (MeV)                           & 
       &     & &( MeV)                   & (MeV)& (MeV)&\\\colrule
$\pi$       &0 &140  &140  &0    &3.63&$\rho$      &0 &775  &775  &769 &1.83\\
$\pi(1300)$ &1 &1300 &1310 &1301 &2.74&$\rho(1450)$&1 &1450 &1571 &1576&1.44\\
$\pi(1800)$ &1 &1812 &1503 &1466 &2.83&$\rho$      &2 &1720 &1946 &2098&1.58\\
\colrule
$K$         &0 &494  &494  &0    &4.13&$K^*$       &0 & 892 &892   &769&1.99\\
$K(1460)$   &1 &1460 &1302 &1301 &1.97&$K^*(1410)$ &1 &1410 &1443 &1576&1.38\\
$K$         &2 &     &1655 &1466 &1.96&$K^*$       &2 &     &1781 &2098&1.44\\
\colrule
$\eta$      &0& 548  & 610  &0   &3.74&$\omega$    &0 &775  &775  &769 &1.83\\
$\eta'$     &0& 958  & 958  &872 &2.73&$\phi$      &0 &1019 &1039 &769 &2.21\\
$\eta(1295)$&1& 1294 & 1138 &1361&2.62&$\phi(1680)$&1 &1680 &1686 &1576&1.55\\
$\eta(1475)$&1& 1476 & 1297 &1516&2.41&$\phi$      &2 &2175 &1897 &2098&1.55\\
\botrule
\end{tabular}}
\label{light_mesons}
\end{table}

The mass spectrum $M_\mathcal{Q}$  of mesons  and quark-meson coupling constants  $h_{\cal Q}$ are determined by the quadratic part of the effective meson action \textit{via} equations
\begin{eqnarray}
\label{mass-eq}
1=
\frac{g^2}{\Lambda^2}\tilde \Pi_{\cal Q}(-M^2_{\cal Q}|B),
\label{eq-spec}
\\
h^{-2}_{\cal Q}=
\frac{d}{dp^2}\tilde\Pi_{\cal Q}(p^2)|_{p^2=-M^2_{\cal Q}}, 
\label{hqm}
\end{eqnarray}
where $\tilde\Pi_{\cal Q}(p^2)$ is the diagonalized two-point correlator $\tilde\Gamma^{(2)}_{\cal QQ'}(p)$  put on mass shell:
\begin{equation*}
\tilde\phi^\dagger_{\mathcal Q}(-p)\left[\mathcal{O}^T(p)\tilde\Gamma^{(2)}(p)\mathcal{O}(p)\right]_{\cal QQ'}\tilde\phi_{{\mathcal Q}'}(p)|_{p^2=-M^2_{\cal Q}}=\tilde\Pi_{\cal Q}(-M_{\mathcal{Q}}^2)\tilde\phi^\dagger_{\mathcal Q}(-p)\tilde\phi_{\mathcal Q}(p)|_{p^2=-M^2_{\cal Q}}.
\end{equation*}
Explicit construction of  $\tilde\Pi_{\cal Q}(p^2)$  will be discussed in the next section.   
Solution of Eq.~\eqref{eq-spec} identifies the position of the pole in the propagator of the meson with quantum numbers ${\cal Q}$.  Definition \eqref{hqm} of the meson-quark coupling constant $h_{\cal Q}$ provides correct residue at the pole.

The free parameters of the model are the IR limits of the running  renormalized strong coupling constant $\alpha_s$, quark masses  $m_{u}=m_{d}$, $m_s$, $m_c$, $m_b$,  and the scales $\Lambda$ and  $ R$.
By construction, the  coupling constant and the quark masses  correspond to the background Feynman gauge condition and momentum subtraction  (MOM) renormalization scheme at subtraction point $p^2=0$ .  The scale  $\Lambda$ and mean domain size  $ R$ are related to the scalar gluon condensate and topological susceptibility of  pure gluodynamics respectively,
\begin{eqnarray*}
\langle\alpha_s F^2\rangle=\frac{2}{3}\frac{\Lambda^4}{\pi}, \quad \chi_\mathrm{YM}=\frac{1}{72}\frac{\Lambda^8R^4}{128\pi^2}. 
\end{eqnarray*}

It should be noted that decomposition \eqref{wavef}  and  \eqref{qmvert} attributes the same radial 
form factor $F_{nl}$ to the mesons  with different  spin-parity $J$.  
 Moreover, the form of $F_{nl}$  appears to be the same for all quarkonium-like collective excitations with different quark content  and spin-parity such as  $\pi$ and $J/\psi$ mesons.  On the contrary, the physical meson states correspond to the momentum dependent transformed basis
 and respectively transformed quark current
\begin{eqnarray*}
f_p^{aJln}(z)=\sum_{n'=0}^{\infty}{\mathcal O}_{aJl}^{nn'}(p)f^{ln'}(z), \ \ 
{\mathcal O}_{aJl}{\mathcal O}_{aJl}^T=I,
\\ 
\tilde J^{aJ}(p,z)=\sum_{nl}^{\infty}\left(z^2\right)^{l/2} f^{aJnl}_{p}(z) \mathcal{O}^{nn'}_{aJl}\tilde J^{aJln'}(p),
\end{eqnarray*}
 where ${{\mathcal O}_{aJl}(p)}$ is an orthogonal transformation of the initial basis taking into account  two-point function  $\tilde\Gamma_{\mathcal{QQ}'}^{(2)}(p)$.  All this means that though \textit{ab initio}  the basic property of quark-meson interaction form factor is set up by gluon propagator, it is the quark loop that  defines its final physical form which  is  different for different mesons.

\section{Masses and decay constants of mesons \label{section_masses_of_mesons}}

\subsection{ Mass spectrum of radial excitations of light, heavy-light mesons and heavy quarkonia }

Meson masses are defined by the algebraic equation \eqref{eq-spec}.   This equation emerges as follows.
In  the momentum representation, the quadratic part of the effective action  pseudoscalar and vector meson fields with zero orbital momentum  has the form
\begin{eqnarray*}
{\cal S}_2=-\frac12\int \frac{d^4p}{(2\pi)^4}
\tilde\Phi^{aV0n}_\mu(-p)
\left[
\Lambda^2\delta^{aa'}\delta_{\mu\mu'}\delta_{nn'}
- g^2
\tilde\Gamma^{(2)\mu\mu'}_{aV0n,a'V0n'}(p)
\right]
\Phi^{a'V0n'}_{\mu'}(p)
\\
\nonumber
-\frac12\int \frac{d^4p}{(2\pi)^4}\tilde\Phi^{aP0n}(-p)\left[\Lambda^2\delta^{aa'}\delta_{nn'}- g^2\tilde\Gamma^{(2)}_{aP0n,a'P0n'}(p^2)\right]\tilde\Phi^{a'P0n'}(p),
\end{eqnarray*}
where  vector two point correlator has the structure
\begin{eqnarray}
\tilde\Gamma_{aV0n,a'V0n'}^{(2)\mu\mu'}(p)=
\tilde\Gamma^{(2)}_{aV0n,a'V0n'}(p^2)\delta_{\mu\mu'}+\tilde L_{aV0n,a'V0n'}(p^2)p_\mu p_{\mu'}.
\label{vector-corr}
\end{eqnarray} 
Vector fields $\phi^{aV0n}$ (see Eq.\eqref{Pphi})  are  subject to the on-shell condition
\begin{eqnarray*}
p^\mu\phi^{aV0n}_\mu =0, \   p^2=-M_{aV0n}^2,
\end{eqnarray*}
while the  mass  $M_{aJ0n}$  ($J=P,V$) is  determined by \eqref{mass-eq} with 
\begin{eqnarray}
\label{Pidiag}
\tilde \Pi_{\cal Q}(p)\longrightarrow \tilde\Pi_{ aJ0}(p)=\mathcal{O}_{aJ0}^{T}(p^2)     
\tilde\Gamma^{(2)}_{aJ0,aJ0}(p^2)
 \mathcal{O}_{aJ0}(p),
\end{eqnarray}
i.e. the diagonalized first term in  Eq.\eqref{vector-corr}.
Only one-loop diagrams (the first diagram in the first line in Fig. \ref{diagrams}) contribute to two-point correlation function $\Pi_{\cal QQ'}$ for all mesons except $\eta$ and $\eta'$. The quadratic part of the effective action and  all other  relations for scalar and axial vector fields can be obtained by the exchange of indices $P\to S$, $V\to A$.

In general, the one loop contribution to $\tilde\Gamma^{(2)}_{aJ0,aJ0}$ in Eq.\eqref{Pidiag} can be expressed in terms of  quark loops  of  the form
\begin{eqnarray}
\tilde\Pi^{nn'}_J\left(-M^2;m_f,m_{f'}\right)=
\frac{\Lambda^2}{4\pi^2}\textrm{Tr}_ v  \int\limits_0^1\! dt_1\! \int\limits_0^1\! dt_2\! \int\limits_0^1\! ds_1\! \int\limits_0^1\! ds_2\! \left(\frac{1-s_1}{1+s_1}\right)^{m_f^2/4 v  \Lambda^2}\left(\frac{1-s_2}{1+s_2}\right)^{m_{f'}^2/4 v  \Lambda^2}\times
\nonumber\\
 t_1^n t_2^{n'}\frac{\partial^n}{\partial t_1^n}\frac{\partial^{n'}}{\partial t_2^{n'}}\frac{1}{\Phi_2^2}\left[\frac{M^2}{\Lambda^2}\frac{F_1^{(J)}}{\Phi_2^2}+\frac{m_f m_{f'}}{\Lambda^2}\frac{F_2^{(J)}}{(1-s_1^2)(1-s_2^2)}+\frac{F_3^{(J)}}{\Phi_2}\right]\exp\left\lbrace \frac{M^2}{2 v  \Lambda^2}\frac{\Phi_1}{\Phi_2}\right\rbrace,
\label{Pi_basic}
\end{eqnarray}
where
\begin{eqnarray}
&& \Phi_1=s_1s_2+2\left(\xi_1^2s_1+ \xi_2^2s_2\right)(t_1+t_2)v,
\nonumber\\
&&\Phi_2=s_1 + s_2 + 2 (1 + s_1 s_2) (t_1 + t_2)  v  + 
 16 (\xi_1^2 s_1 + \xi_2^2 s_2) t_1 t_2  v^2,
 \nonumber\\
&& F_1^{(P)}=(1+s_1s_2)\left[2(\xi_1s_1+\xi_2s_2)(t_1+t_2) v +4\xi_1\xi_2(1+s_1s_2)(t_1+t_2)^2 v^2+ s_1s_2(1-16\xi_1\xi_2t_1t_2 v^2)\right],
\nonumber\\
&& F_1^{(V)}=\left(1-\frac13s_1s_2\right)\left[s_1s_2+16 \xi_1 \xi_2 t_1 t_2  v^2+2(\xi_1s_1+\xi_2s_2)(t_1+t_2) v \right]+
 4\xi_1\xi_2 (1-s_1^2s_2^2)(t_1-t_2)^2 v^2,
\nonumber \\
&& F_2^{(P)}= (1 + s_1 s_2)^2,\quad F_2^{(V)}= (1 - s_1^2 s_2^2),
\nonumber\\
&& F_3^{(P)}=4  v (1 + s_1 s_2) (1 - 16 \xi_1 \xi_2 t_1 t_2  v^2),\ F_3^{(V)}=2  v (1 - s_1 s_2) (1 - 16 \xi_1 \xi_2 t_1 t_2  v^2),
\nonumber\\
&& F_1^{(S)}=F_1^{(P)}, \ F_1^{(A)}=F_1^{(V)}, 
\nonumber\\
&& F_2^{(S)}=-F_2^{(P)}, \ F_2^{(A)}=-F_2^{(V)},
\label{SA}
\\
&& F_3^{(S)}=F_3^{(P)}, \ F_3^{(A)}=F_3^{(V)}.
\nonumber
\end{eqnarray}
After diagonalization with respect to $(n,n')$, function \eqref{Pi_basic} contains information about the masses of all radial excitations of  light, heavy-light  mesons and heavy quarkonia  with $J=S,P,V,A$  and zero orbital momentum  $l=0$.  
One can see from Eq.\eqref{SA} that expressions for $\tilde\Pi^{nn'}_J$ with the same spin but opposite parity differ only in the sign of the function $F_2$ (second term in square brackets in \eqref{Pi_basic}).  	  This difference has a  peculiar consequence  for meson spectrum. Real $M^2$ solutions of Eq.\eqref{eq-spec} for both scalar and axial mesons are absent,  while pseudoscalar and vector meson solutions exist irrespective to the quark content of a meson.  In the present approach scalar and axial mesons as quark-antiquark collective excitations analogous to the corresponding pseudoscalar and vector mesons are absent in the spectrum.  However, scalar and axial mesons naturally appear in the hyperfine splitting structures of the orbital excitations of  vector mesons~\cite{EN}. For example, $\sigma$-meson as a plain  analogue of $\pi$-meson is absent. The reason is that the term in Eq.\eqref{Pi_basic} proportional to the quark masses dominates both in the case of heavy and light quarks. For heavy quarks it dominates just because of their large masses. For the light quarks, due to the contribution of  zero  modes to the scalar part of quark propagator \eqref{qprop}  this term  dominates  again. 
As a result, solutions to Eq.\eqref{eq-spec}	for scalar and axial states are absent in the whole range of the quark masses.
Further below we will not discuss scalar and axial mesons any more.  The study of the spectrum of parity partners  in a more detailed and systematic way than the estimates of paper~\cite{EN} has to be done. It will be presented elsewhere.

\begin{table}[ph]
\caption{Masses of heavy-light mesons and their lowest radial excitations.}
{\begin{tabular}{@{}ccccc|ccccc@{}} \toprule
Meson & $n$ & $M_{\rm exp}$\cite{PDG} & $M$ &$h$& Meson & $n$ & $M_{\rm exp}$\cite{PDG} & $M$&$h$\\
&&MeV& MeV&&&&MeV& MeV\\
\colrule
$D$  &0 &1864                       &1715 &5.93 &$D^*$  &0 &2010                      &1944&2.94\\
$D$  &1 &                           &2274 &2.56 &$D^*$  &1 &                          &2341&1.74\\
$D$  &2 &                           &2508 &2.32 &$D^*$  &2 &                          &2564&1.66\\\colrule
$D_s$&0 &1968                       &1827 &6.94 &$D_s^*$&0 &2112                      &2092&3.3 \\
$D_s$&1 &                           &2521 &2.53 &$D_s^*$&1 &                          &2578&1.75 \\
$D_s$&2 &                           &2808 &2.42 &$D_s^*$&2 &                          &2859&1.72  \\\colrule
$B$  &0 &5279                       &5041 &9.15 &$B^*$  &0 &5325                      &5215&4.82 \\
$B$  &1 &                           &5535 &3.9  &$B^*$  &1 &                          &5578&2.88  \\
$B$  &2 &                           &5746 &3.4  &$B^*$  &2 &                          &5781&2.4 \\\colrule
$B_s$&0 &5366                       &5135 &10.73&$B_s^*$&0 &5415                      &5355&5.39 \\
$B_s$&1 &                           &5746 &3.75 &$B_s^*$&1 &                          &5783&2.54  \\
$B_s$&2 &                           &5988 &3.42 &$B_s^*$&2 &                          &6021&2.23  \\\colrule
$B_c$&0 &6277                       &5952 &14.86&$B_c^*$&0 &6314 \cite{Dowdall:2012ab}&6310&7.61 \\
$B_c$&1 &6842 \cite{Dolezal:2015gpa}&6904 &3.87 &$B_c^*$&1 &6905 \cite{Dowdall:2012ab}&6938&2.81 \\
$B_c$&2 &                           &7233 &4    &$B_c^*$&2 &                          &7260&2.76 \\
 \botrule
\end{tabular} }
\label{heavy-light}
\end{table}

Two-point correlators for $\eta^0$ and $\eta^8$  include  additional contribution  described by the two-loop diagram in the first line of  Fig.\ref{diagrams}.   This additional contribution contains two tadpole diagrams integrated over all configurations of the background field.  The tadpole diagram has the form
\begin{gather}
G^{(1)}_{aPn}=\mathrm{Tr}\lambda^a i\gamma_5 F_{0n}(x|B)S(x,x|B)=\pm i\frac{\Lambda^3}{2\pi^2}\sum_f \lambda^a_{ff}R^n_f,
\nonumber\\
R_f^n=\mathrm{Tr}_v \frac{v m_f}{\Lambda}  \int_0^1 dt t^n\int_0^1 ds  \frac{\partial^n}{\partial t^n} \frac{1}{(2vt+s)^2} \left(\frac{1-s}{1+s}\right)^{m_f^2/4v\Lambda^2} \frac{s^2}{1-s^2},
\label{Rfn}
\end{gather} 
where  sign "$\pm$"  corresponds to the self- and anti-selfdual background fields.  Two-point correlator  in momentum space reads
\begin{equation*}
\Gamma_{ab}^{(2)nn'}(p^2)=\Pi^{nn'}_{ab}(p^2)-\delta \Pi^{nn'}_{ab}(p^2),
\end{equation*}
where $\Pi(p^2)$ is the one-loop contribution  expressed in terms of  functions $\tilde\Pi^{nn'}_P$  (see Eq.~\eqref{Pi_basic}),  and $\delta\Pi(p^2)$ is a contribution of the two-loop diagram in Fig.\ref{diagrams}, 
\begin{equation}
\label{two_loop_contribution}
\delta \Pi^{nn'}_{ab}(p^2)=\frac{32}{3\pi^4}\Lambda^2 (\Lambda R)^4 \sum_{ff'}\lambda^a_{ff}\lambda^b_{f'f'} R_f R_{f'}\tilde \Xi_2(p^2).
\end{equation}
Here $\tilde \Xi_2$ is the momentum representation of the two-point correlator of the background field $B$ in the spherical domain approximation~\cite{NK4}
\begin{equation*}
\tilde \Xi_2(p^2)=\int_0^1 dt \sqrt{1-t^2} \int_{0}^{1} ds\ s \cos\left(\sqrt{4p^2R^2t^2s}\right)
\left(\frac{3\pi}{2}-3\arcsin{\sqrt{s}}-(5-2s)\sqrt{s(1-s)}\right).
\end{equation*}
Solving  Eq.\eqref{mass-eq} with completely (i.e. over radial and flavour indices) diagonalized correlator 
one finds masses of $\eta,\eta'$ and their excited states.

The values of parameters given in Table~\ref{values_of_parameters} were fitted to the ground state of $\pi $, $\rho $, $K$, $K^*$, $ J/\psi $, $\Upsilon $ and $\eta'$ mesons.  The fit can be successfully done irrespective to the number $N$ of radially excited states used for diagonalization of the quadratic part of the action. However, the fitted values of parameters depend on $N$. Figure~\ref{deltaPN}
illustrates dependence of relative variation of the model parameters on $N$. The iterations converge with $N$ for all parameters faster than $\Delta_N\approx \Delta_1\exp(-.35N)$ for $N>5$.
For $N=7$  the variations of light quark mass $m_{u/d}$ and coupling constant $g$ slow down to one percent level,  while the change of the scale $\Lambda$ and the strange quark mass $m_s$ approach a  fraction of a percent.   Parameters given in Table~\ref{values_of_parameters} and used for calculation of all masses and decay constants  correspond to $N=7$.

\begin{table}[ph]
\caption{Masses of heavy quarkonia.}
{\begin{tabular}{@{}ccccc@{}} \toprule
Meson&$n$&$M_{\rm exp}$\cite{PDG}&$M$&$h$\\
&&(MeV) & (MeV)&\\
\colrule
$\eta_c(1S)$  &0 &2981 &2751&9.95\\
$\eta_c(2S)$  &1 &3639 &3620&3.45\\
$\eta_c$      &2 &     &3882&3.29\\
\colrule
$J/\psi(1S)$  &0 &3097 &3097&4.87\\
$\psi(2S)$    &1 &3686 &3665&2.12\\
$\psi(3770)$  &2 &3773 &3810&2.27\\
\colrule
$\Upsilon(1S)$&0 &9460 &9460 &10.6\\
$\Upsilon(2S)$&1 &10023&10102&3.94\\
$\Upsilon(3S)$&2 &10355&10249&2.48\\\botrule
\end{tabular} \label{ta1}}
\label{heavy}
\end{table}

The results of computation of the masses of light mesons and their lowest radial excitations are given in Table~\ref{light_mesons}. The rightmost column  demonstrates behaviour of meson masses in the chiral limit  as it has been defined in \cite{NK4}. Since the quark masses here have the meaning of IR limit of the running effective mass,  the  appropriate way to turn the system into the chiral limit is to alter the  masses of quarks $m_{u/d}$ and $m_s$  to
the value $\tilde m$ 
\begin{equation}
\label{chiral}
\tilde{m}_{u/d}=\tilde{m}_s=\tilde{m}=136\ \text{MeV},
\end{equation}
at which  the light pseudoscalar octet mesons become massless.  Then the 
current quark masses $\mu_f$ may be found as differences
\begin{gather*}
\mu_{u/d}=m_{u/d}-\tilde{m}=9 \ \mathrm{ MeV}, \ 
\mu_s=m_s-\tilde{m}=240\ \mathrm{ MeV}.
\end{gather*}
Unlike the current masses themselves, their ratio is renormalization group invariant.  The ratio takes the value
\begin{equation*}
\frac{\mu_s}{\mu_{u/d}}=26.7,
\end{equation*}
 that is close to the generally recognized value  and just slightly differs from the result of \cite{NK4} where diagonalization has been ignored.

It follows from Eq.\eqref{Rfn}  that  in the chiral limit \eqref{chiral} a degeneracy emerges,
 $$R_{u/d}(p^2)=R_{s}(p^2),$$
and according to Eq.\eqref{two_loop_contribution} mixing between  $\eta^0$ and $\eta^8$ disappears. The two-loop diagram contributes only to the correlator of $\eta^0$. 
As a result,  $\eta$ meson becomes massless simultaneously with pions and kaons, but the $\eta'$ meson stays massive with a slightly reduced mass. This mechanism provides resolution of the $U_A(1)$ problem as it is seen  in terms of $\eta'$ mass. A basic scheme of simultaneous resolution  of the  $U_A(1)$  and the strong CP-problem in terms of the quark eigenmodes  was elaborated in papers~\cite{NK4,NK6} within the spherical domain approximation.

Results of numerical calculation of the masses of  ground state and two first radial excitations of light, heavy-light mesons and heavy quarkonia are given in Tables~\ref{light_mesons}, \ref{heavy-light} and \ref{heavy}. Overall inaccuracy of description is  less than 15\%  besides the second radial pion excitation  $\pi(1800)$ where it rises to 17\%.  It has to be stressed that there are rigid asymptotic regimes which drive the three regions of meson spectrum~\cite{EN,EN1}: chiral symmetry breaking and dynamical quark confinement for the light mesons,  proper Isgur-Wise limit for the case of heavy-light mesons  and correct 
UV behaviour of the gluon and quark propagators combined with the dynamical quark confinement for the heavy quarkonia.

\subsection{$V\to \gamma$ transition constants 
\label{v_gamma_transition constants}}

\begin{figure}\centering
\includegraphics[scale=1.2]{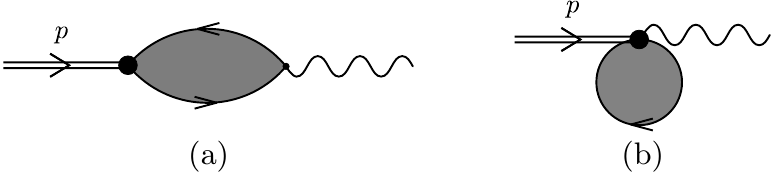}
\caption{Two diagrams contributing to 
$g_{V\gamma}$.\label{g_rho_gamma_diagrams}}
\end{figure}

The amplitude of vector meson decay into a leptonic pair is given by the 
formula
\begin{equation*}
A_{V(p) \to \bar{l}(q)l(p+q)}=e^\mu(p) \mathcal{M}^{\mu\nu} 
\bar{l}(q)\gamma^\nu l(p+q),
\end{equation*}
where $e^\mu$ is polarization vector of a meson.
Two diagrams contributing to $\mathcal{M}_{\mu\nu}$
\begin{gather*}
\mathcal{M}_{\mu\nu}(p) = \mathcal{M}_{\mu\nu}^\mathrm{(a)}(p) + 
\mathcal{M}_{\mu\nu}^\mathrm{(b)}(p)  = C h_V \left(\left[ 
I^\mathrm{(a)}_\bot(p^2) + I^\mathrm{(b)}_\bot (p^2)
\right]\left(\delta_{\mu\nu}p^2-p_\mu p_\nu\right) + \left[ 
I^\mathrm{(a)}_\| (p^2)+ I^\mathrm{(b)}_\|(p^2) \right]p_\mu p_\nu\right),\\
g_{V\gamma} = C h_V \left[ I^\mathrm{(a)}_\bot (-M_V^2)+ I^\mathrm{(b)}_\bot (-M_V^2)
\right],
\end{gather*}
are shown in Fig.\ref{g_rho_gamma_diagrams}. Constant $C$ originates 
from the flavour  content of a meson and  quark charges:
\begin{center}
\begin{tabular}{cccccc}
Meson&$\rho$ &$\omega$ &$\phi$ &$J/\psi$ &$\Upsilon$\\
  C&\hspace*{0.5em}$1/\sqrt{2}$\hspace*{0.5em}&\hspace*{0.5em} 
$1/3\sqrt{2}$\hspace*{0.5em}& \hspace*{0.5em}$1/3$\hspace*{0.5em}& 
\hspace*{0.5em}$2/3$\hspace*{0.5em}&\hspace*{0.5em} $1/3$\hspace*{0.5em}
\end{tabular}
\end{center}
Exact form of vertex 
operator is not important for gauge invariance, as it can be seen from Eq.\eqref{1-photon-vertex}. Hence, we can use regularization
\begin{equation*}
F_{n0}^\varepsilon = \int_\varepsilon^1 dt\ t^n 
\frac{\partial^n}{\partial t^n} \exp\left[ t \left( 
\frac{\stackrel{\leftrightarrow}{\cal D}}{\Lambda}\right)^2\right].
\end{equation*}
Properly regularized contribution of the first diagram,  Fig.\ref{g_rho_gamma_diagrams}a, is
\begin{gather*}
\begin{split}
\mathcal{M}_{\mu\nu}^\mathrm{(a)} &  = Ch_{V}\sum_{n'} \mathcal{O}_{n'0}(p^2)\int 
d\sigma \int \frac{d^4p'}{2\pi^4}\int d^4x e^{ipx}\int d^4y e^{ip'y} 
\mathrm{Tr}S(y,x)V^{n'}(x)S(x,y)\gamma_\mu \\
& =
Ch_{V}\sum_{n'} \mathcal{O}_{n'0}(p^2)\mathrm{Tr}_v\frac{1}{16 \pi^2} \int_0^1 
ds_1 \int_0^1 ds_2 \int_\varepsilon^1  dt 
\left(\frac{1-s_1}{1+s_1}\right)^{m_f^2/4v\Lambda^2} 
\left(\frac{1-s_2}{1+s_2}\right)^{m_f^2/4v\Lambda^2} \\
& \times t^{n'}\frac{\partial^{n'}}{\partial 
t^{n'}}\frac{1}{\Phi_1^2}\left[\delta_{\mu\nu}\Phi_2-p_\mu 
p_\nu\Phi_3\right]\exp\left(-\frac{p^2}{4v\Lambda^2} \Phi_4\right),
\end{split}\\
\Phi_1=s_1+s_2+2(1+s_1s_2)tv,\\
\Phi_2=-\frac{4m_q^2 (1 - s_1^2s_2^2)}{ (1 - s_1^2)(1 - s_2^2)}
+\frac{4p^2 (s_1 s_2 (3 - s_1 s_2) + (s_1 + s_2)(3 -s_1 s_2) t v+ 3(1 - 
s_1^2 s_2^2) t^2 v^2)}{3 (s_1 + s_2 + 2(1 + s_1 s_2) t v)^2}
-\frac{8v(1- s_1^2 s_2^2)}{s_1 + s_2 + 2(1 + s_1 s_2) t v},\\
\Phi_3=\frac{8 s_1 s_2 (3 + s_1 s_2) + 8 (s_1 + s_2) (3 + s_1 s_2) t v +
  24 (1 - s_1^2 s_2^2) t^2 v^2}{3 (s_1 + s_2 + 2(1 + s_1 s_2) t v)^2},\quad
\Phi_4=\frac{2 s_1 s_2 + (s_1 + s_2) t v}{s_1 + s_2 + 2 (1 + s_1 s_2) t v}.
\end{gather*}

Contribution of  the diagram  shown at  Fig.\ref{g_rho_gamma_diagrams}b looks as
\begin{gather*}
\begin{split}
\mathcal{M}_{\mu\nu}^\mathrm{(b)} &  =Ch_{V}\sum_{n'} \mathcal{O}_{n'0}(p^2)\int 
d\sigma \mathrm{Tr}S(x,x)V^{n'}_A(x)\gamma_\mu \\
& =
Ch_{V}\sum_{n'} \mathcal{O}_{n'0}(p^2) \mathrm{Tr}_v\frac{v}{8 \pi^2} \int_0^1 ds 
\int_0^1 d\tau \int_\varepsilon^1 dt 
\left(\frac{1-s}{1+s}\right)^{m_f^2/4v\Lambda^2} 
t^{n'}\frac{\partial^{n'}}{\partial t^{n'}} 
\frac{1}{\Phi_5^2}\left[\delta_{\mu\nu}\Phi_6-p_\mu 
p_\nu\Phi_7\right]\exp\left(-\frac{p^2}{4v} \Phi_8\right),
\end{split}\\
\Phi_5=s+2tv,\quad
\Phi_6=\frac{8tv}{s+2tv}=2-\frac{2s-4tv}{s+2tv},\quad
\Phi_7=4 s t^2 \tau^2 v,\quad
\Phi_8=\frac{s t \tau^2 v}{s+2tv}.
\end{gather*}
Form factors $I^\mathrm{(a)}_\bot$ and $I^\mathrm{(b)}_\bot$ contain divergences that 
cancel each other, so $I^\mathrm{(a)}_\bot+I^\mathrm{(b)}_\bot$ is 
finite after removal of regularization  $\varepsilon\to 0$. Let us 
demonstrate this for ground state $n=0$:
\begin{multline}\label{div_cancellation}
I^\mathrm{(a)}_\bot+I^\mathrm{(b)}_\bot=
\left(I^\mathrm{(a)}_\bot+\mathrm{Tr}_v\frac{1}{16 \pi^2} \int_0^1 ds_1 
\int_0^1 ds_2 \int_\varepsilon^1  dt \frac{8v}{(s_1 + s_2  + 
2vt_1)^3}\right)+\\
\left(I^\mathrm{(b)}_\bot-\mathrm{Tr}_v\frac{v}{8 \pi^2} \int_0^1 ds_1 
\int_0^1 ds_2 \int_\varepsilon^1  dt \left[\frac{2}{(s + 
2vt)^2}-\frac{2s-4vt}{(s+2vt)^3}\right]\right)+\\
\mathrm{Tr}_v\left[-\frac{1}{16 \pi^2} \int_0^1 ds_1 \int_0^1 ds_2 
\int_\varepsilon^1  dt \frac{8v}{(s_1 + s_2  + 2vt_1)^3}  + \frac{v}{8 
\pi^2} \int_0^1 ds_1 \int_0^1 ds_2 \int_\varepsilon^1  dt \frac{2}{(s + 
2vt)^2}\right]\\
-\mathrm{Tr}_v\frac{v}{8 \pi^2} \int_0^1 ds_1 \int_0^1 ds_2 
\int_\varepsilon^1  dt \frac{2s-4vt}{(s+2vt)^3}.
\end{multline}
Terms in round parentheses are finite in the limit $\varepsilon \to 0$. 
Terms in square brackets read
\begin{multline*}
-\frac{1}{16 \pi^2} \int_0^1 ds_1 \int_0^1 ds_2 \int_\varepsilon^1 dt 
\frac{8v}{(s_1 + s_2  + 2vt_1)^3}  + \frac{v}{8 \pi^2} \int_0^1 ds_1 
\int_0^1 ds_2 \int_\varepsilon^1  dt \frac{2}{(s + 2vt)^2}=\\
\frac{v}{4\pi^2}\left(\frac{1}{2v}\ln\frac{\varepsilon (1+2v)^2 (1 + 
\varepsilon v)}{(1+v)(1+ 2\varepsilon v)^2 }- \frac{1}{2v}\ln 
\frac{\varepsilon (1+2v)}{1+ 2\varepsilon v}\right)=\\
\frac{1}{8\pi^2} \ln \frac{(1+2v) (1+\varepsilon 
v)}{(1+v)(1+2\varepsilon v)} \xrightarrow{\varepsilon \to 0} \frac{1}{8 
\pi^2} \ln \frac{1+2v}{1+v}.
\end{multline*}
The limit $\varepsilon \to 0$ of the  last term in Eq.\eqref{div_cancellation} reads
\begin{equation*}
-\mathrm{Tr}_v\frac{v}{8 \pi^2} \int_0^1 ds_1 \int_0^1 ds_2 
\int_\varepsilon^1  dt \frac{2s-4vt}{(s+2vt)^3} = 
\frac{v}{4\pi^2}\frac{1-\varepsilon}{(1+2v)(1+2\varepsilon 
v)}\xrightarrow{\varepsilon \to 0} \frac{v}{4\pi^2} \frac{1}{1+2v}.
\end{equation*}
Gauge invariance requirement
\begin{equation*}
I^\mathrm{(a)}_\|+I^\mathrm{(b)}_\|=0
\end{equation*}
holds, which has been checked numerically.

\begin{table}[ph]
\caption{Decay and transition constants of various  mesons}
{\begin{tabular}{@{}cccc|cccc@{}} \toprule
Meson&$n$&$f_P^{\rm exp}$&$f_P$&Meson&$n$&$g_{V\gamma}$ 
\cite{PDG}&$g_{V\gamma}$\\
&&(MeV)& (MeV)&&&\\
\colrule
$\pi$      &0 &130 \cite{PDG}        &140 & $\rho$&0&0.2&0.2 \\
$\pi(1300)$&1 &                      &29 & $\rho$&1&&0.053 \\
\colrule
$K$        &0 &156 \cite{PDG}        &175  & $\omega$&0&0.059&0.067\\
$K(1460)$  &1 &                      &27   & $\omega$&1&&0.018\\
\colrule
$D$        &0 &205 \cite{PDG}        &212  & $\phi$&0&0.074&0.071\\
$D$        &1 &                      &51   & $\phi$&1&&0.02\\
\colrule
$D_s$      &0 &258 \cite{PDG}        &274  & $J/\psi$&0&0.09&0.06\\
$D_s$      &1 &                      &57  & $J/\psi$&1&&0.015\\
\colrule
$B$        &0 &191 \cite{PDG}        &187  & $\Upsilon$&0&0.025&0.014\\
$B$        &1 &                      &55   & $\Upsilon$&1&&0.0019\\
\colrule
$B_s$      &0 &253 \cite{Chiu:2007bc}&248  &  & &\\
$B_s$      &1 &                      &68   &  & &\\
\colrule
$B_c$      &0 &489 \cite{Chiu:2007bc}&434  &  & &\\
$B_c$      &1 &                  &135  & &&\\
\botrule
\end{tabular}
\label{constants}}
\end{table}

Numerical values of transition constants are given in 
Table~\ref{constants}. Though the masses of $\rho$ and $\omega$ mesons are equal 
to each other,  their transition constants  $g_{V\gamma}$   differ  due to  isospin.  Transition constants $g_{V\gamma}$ for heavy quarkonia turn out to be underestimated. Though  a clear reason for this has not been identified yet,  it could be due to the necessity to take into account larger $N$ in calculations related to heavy quarkonia.

\subsection{Leptonic decay constants \label{section_leptonic_decay_constants}}
Leptonic decay constant is defined as
\begin{gather*}
M(P_n\rightarrow l \overline\nu)=i\frac{G_F}{\sqrt{2}}\mathcal{K}F_n\left(p^2\right) \Phi_P(k)k_\mu \overline{l}(k')\gamma_\mu(1-\gamma_5)\nu(k+k'),\\
f_{P_n}=F_n\left(-M^2_n\right),
\end{gather*}
where $\mathcal{K}$ is CKM matrix element corresponding to a given meson.

\begin{figure}\centering
\includegraphics[scale=1.2]{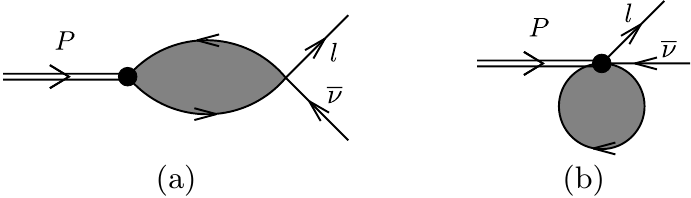}
\caption{Two diagrams contributing to $f_P$. \label{weak_decay_diagrams}}
\end{figure}

The contributions to $F_n$ of diagrams (a) and (b)  shown in Fig. \ref{weak_decay_diagrams} are given by the formulas
\begin{gather*}
F^{(a)}_n(p^2)=h_{P_n}\sum_{n'} \mathcal{O}_{n'n}(p^2)\int d\sigma \int \frac{d^4p'}{2\pi^4}\int d^4x e^{ipx}\int d^4y e^{ip'y} \mathrm{Tr}S_f(y,x)V^{n'}(x)S_{f'}(x,y)\gamma_\mu (1-\gamma_5)\\
F^{(b)}_n(p^2)=h_{P_n}\sum_{n'} \mathcal{O}_{n'n}(p^2)\int d\sigma  \frac{1}{\mathcal{K}} \sum_f\mathrm{Tr}S_f(x,x)V^{n'}_{Wff}(x)
\end{gather*}
$\mathcal{O}_{nn'}$ is the matrix that diagonalizes polarization operator $\tilde\Pi^{nn'}_P$ corresponding to  meson under consideration.
After standard calculations  one obtains the following expression for $f_{P_n}$
\begin{eqnarray}
&&f_{P_n}=h_{P_n}\sum_{n'} \mathcal{O}_{n'n}(-M_n^2)\frac{1}{4\pi^2} \left\{\mathrm{Tr}_v \int_0^1\!\int_0^1\!\int_0^1\! dt ds_1 ds_2 \left(\frac{1-s_1}{1+s_1}\right)^{m_{f_1}^2/4v\Lambda^2}\left(\frac{1-s_2}{1+s_2}\right)^{m_{f_2}^2/4v\Lambda^2} \right.
\nonumber\\
&&\times t^{n'}\frac{\partial^{n'}}{\partial t^{n'}}\frac
{1+s_1s_2}{(s_1 + s_2+2(1+ s_1s_2)tv)^3}\left(m_{f_1}\frac{s_1+2tv\left(1-\xi_1\left(1+ s_1^2\right)\right)}{1 -s_1^2} +m_{f_2}\frac{s_2 + 2tv\left( 1-\xi_2\left(1+s_2^2\right) \right)}{1-s_2^2}\right)
\nonumber\\
&&\times \exp\left({\frac{M_{Pn}^2}{2v\Lambda^2}\frac{s_1s_2+2\left(\xi_1^2 s_1+\xi_2^2s_2\right)tv}{s_1+s_2+2(1+s_1s_2)tv}}\right)-
\nonumber\\
 &&- 2\xi_1 m_{f_1}\mathrm{Tr}_v \int_0^1 \int_0^1 \int_0^1 ds dt d\tau \left(\frac{1-s}{1+s}\right)^{m_{f_1}^2/4v\Lambda^2} t^{n'} \frac{\partial^{n'}}{\partial t^{n'}} \frac{vst\tau}{(s+2vt)^3}\exp \left(\frac{M_{Pn}^2}{\Lambda^2}\frac{\xi_1^2 st\tau^2}{s+2vt}\right) 
\nonumber\\
&&\left.-2\xi_2 m_{f_2}\mathrm{Tr}_v\int_0^1 \int_0^1 \int_0^1 ds dt d\tau \left(\frac{1-s}{1+s}\right)^{m_{f_2}^2/4v\Lambda^2} t^{n'}\frac{\partial^{n'}}{\partial t^{n'}} \frac{ vst\tau}{(s+2vt)^3}\exp \left( \frac{M_{Pn}^2}{\Lambda^2}\frac{\xi_2^2 st\tau^2}{s+2vt}\right)\right\} .
\label{fpexp}
\end{eqnarray}

Numerical values of several leptonic decay constants are given in Table~\ref{constants}. In agreement with general expectations based on arguments related to the chiral symmetry breaking~\cite{Holl:2004fr} and finite energy sum rules~\cite{krasnikov,kataev}, decay constants are order of magnitude smaller  for excited states than for  the ground state mesons. In the present approach this sharp decrease is a highly nontrivial feature  since the integrand in \eqref{fpexp} includes exponents of $M_{Pn}^2/\Lambda^2\sim  10$, and naively one would expect large decay constants. However, the  chiral symmetry realisation  combined with the   orthogonal transformation $\mathcal{O}$  correctly leads to  very small value.

\section{Discussion}
\label{discussion}

In this section we  touch on  several  issues  which have not been fully elaborated yet but appear to be very important  as they allow one to  identify the place of the  present approach  among  other   models  of confinement, chiral symmetry breakdown  and hadronization as well as the methods underlining them.   These are potential  interrelations of the present approach to the soft wall AdS/QCD models, comparison of the properties of gluon correlator \eqref{gprop}  with the  Landau gauge gluon and quark propagators as they appear in functional renormalization group, Schwinger-Dyson Equations and lattice QCD, as well as  details of  chiral symmetry realization in the present approach.  
Some basic properties of these  approaches  seem to be visible from the viewpoint of the present formalism.

\subsection{AdS/QCD and harmonic confinement}

Bosonization of the four-quark interaction \eqref{4ferm-1} in terms of bilocal  meson-like fields
$\Phi_{Jc}(x,z)$ leads to the following quadratic part of the effective action
\begin{eqnarray}
&&\mathcal{S}_2=-\frac{1}{2}\int d^4x\int d^4z D(z)\Phi^2_{Jc}(x,z) 
\nonumber\\
&&-
2g^2\int d^4x d^4x' d^4z d^4z'  D(z)D(z')  
\Phi_{Jc}(x,z)\Pi_{Jc,J'c'}(x,x';z,z')\Phi_{J'c'}(x',z'),
\nonumber\\ 
&&\Pi_{Jc,J'c'}(x,x';z,z')={\rm Tr} V_{Jc}(x,z) S(x,x') V_{J'c'}(x',z') S(x',x),
\nonumber\\
&& V_{Jc}(x,z)=\Gamma_J t_c\exp\left\{ iz_\mu  \stackrel{\leftrightarrow}{\mathcal{D}}_\mu(x) \right\}.
\label{bilocm}
\end{eqnarray}
 Meson eigenfunctions $f^{nl}_{\mu_1\dots \mu_l}(z)$ in the decomposition 
$$
\Phi^{aJ}(x,z)=\sum_{nl}\left(z^2\right)^{l/2} f^{nl}_{\mu_1\dots \mu_l}(z) \Phi^{aJln}_{\mu_1\dots \mu_l}(x)
$$
are defined by the action $S_2$  \textit{via } corresponding integral equation. Solution of this  eigenfunction  problem is equivalent to diagonalization of the quadratic part  of the effective action in Eq.\eqref{effective_meson_action}. Specific Gaussian form \eqref{glprop-1} of gluon propagator $D(z)$ is the reason for the radial part of the wave functions to be represented in terms of the generalized Laguerre polynomials, see Eq.~\eqref{wavef}. The mass spectrum has Regge character~\cite{EN}.

For the quadratic in $z$ dilaton profile $\varphi(z)=\kappa z^2$ soft wall AdS/QCD models arrive at the decomposition
\begin{eqnarray*}
\Phi_j(x,z)=\sum_n \phi_{nj}(z)\Phi_{nJ}(x)
\end{eqnarray*}
with the radial meson wave functions proportional to generalized Laguerre polynomials,
$$
\phi_{nj}=R^{j-3/2}\kappa^{1+l}z^{l-j+2} L^l_n(\kappa^2z^2),
$$
which is a solution of the  eigenfunction problem in  differential form. The eigenvalues can be treated as  the meson masses squared, and they  strictly correspond  to Regge spectrum for the quadratic in $z$ dilaton field.

 There are obvious differences between effective action \eqref{bilocm} and the soft wall AdS/QCD action. 
The fifth space-time coordinate $z$ of AdS/QCD model appears   in the present approach  as a distance between quark and anti-quark.   There are four  $z$-coordinates in  \eqref{bilocm}, and  hence the meson wave function contains angular part. Also it is not immediately clear where AdS metrics could come from in   Eq.\eqref{bilocm}. Apart from this, the feature in common is  the Gaussian  weight function which  in both cases plays the most important role for Regge character of the mass spectrum.  
After integrating out  $z$ coordinate the quadratic part of the effective meson action put on-shell (that is neglecting terms of order $(p^2-M^2)^k$ with $k\ge 2$),
\begin{eqnarray*}
&&\mathcal{S}_2=\frac{1}{2}\sum_{aJln}\int d^4x\tilde\phi_{aJln}(-p)\left(p^2-M^2_{aJln}\right) \tilde\phi_{aJln}(p),
\end{eqnarray*}
is equivalent to the effective actions constructed within the AdS/QCD models.

 One gets an impression that the form of dilaton profile in soft wall AdS/QCD approach could be  linked to the gluon propagator and thus to the properties of QCD vacuum.  
A systematic verification of this conjecture requires derivation of the approximate (or reduced to some appropriate limit) differential form of the action \eqref{bilocm}.
Weight and basis functions of the domain model have to be systematically compared with those employed in  soft-wall AdS/QCD \cite{Karch:2006pv} and harmonic confinement \cite{Leutwyler:1977vz,Leutwyler:1977vy, Leutwyler:1977pv, Leutwyler:1977cs, Leutwyler:1978uk} approaches.

\begin{figure}
\begin{centering}
\includegraphics[width=0.3\textwidth]{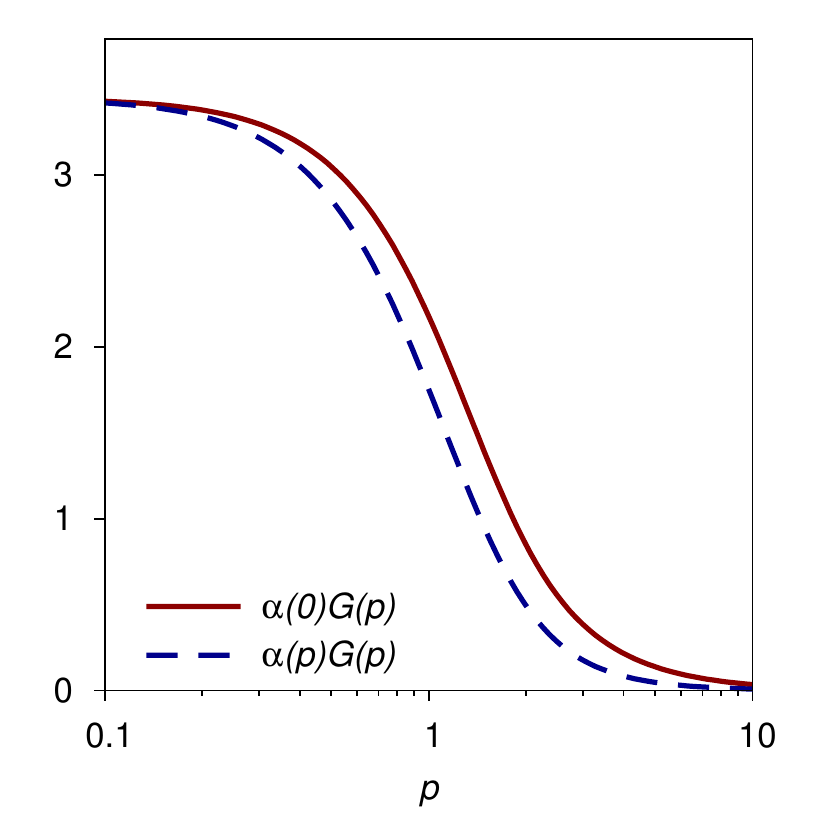}\includegraphics[width=0.3\textwidth]{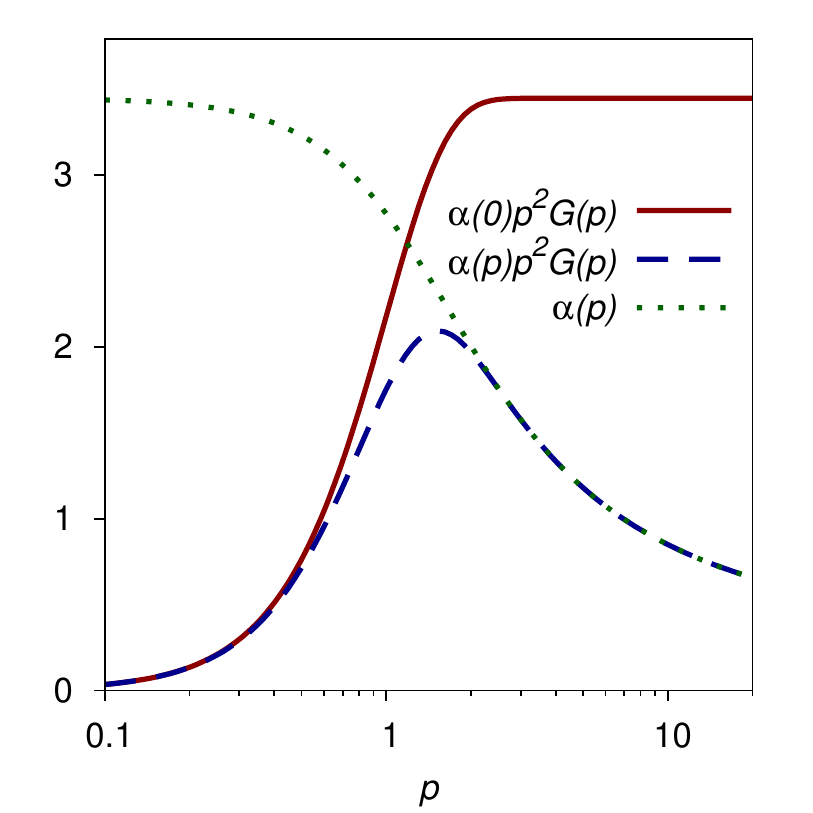}
\end{centering}
\caption{Momentum dependence of the gluon propagator (LHS plot) and dressing function (RHS plot) without (solid line)  and with (dashed line) accounting for the running of the strong coupling constant $\alpha_s(p)$  (dotted line) modelled by the function \eqref{alphasp}. The dashed line reproduces the shape of the Landau gauge dressing function of gluons calculated within functional renormalization group~\cite{Jan2015} and  Lattice QCD~\cite{Bowman, Ilgenfritz} as well as a part of the input gluon propagator used in the approaches based on combined  Dyson-Schwinger and Bethe-Salpeter equations~\cite{Maris,Fischer:2014xha,Dorkin:2014lxa}.  \label{gluon_fig}}
\end{figure}

\subsection{Gluon propagator: Landau gauge DSE, FRG, LQCD vs Abelian (anti-)self-dial field}

It is interesting to take a look at the properties of gluon \eqref{gprop} and quark \eqref{qprop} correlators  in view of the known functional form of quark  and Landau gauge gluon propagators calculated within the functional renormalization group, Lattice QCD and Dyson-Schwinger equations \cite{Bowman,Ilgenfritz,Jan2015,Maris,Fischer:2014xha,Dorkin:2014lxa}.  We do not intend to compare the propagators on a detailed quantitative basis  if for no other reason than difference in gauge condition.
Unlike just mentioned studies, the present approach have assumed background gauge condition so as renormalization of quark-gluon coupling constant is related to the gluon field renormalization. 
The tree level gluon propagator \eqref{gprop} without   radiative corrections  multiplied by the IR limit of the running
coupling constant $\alpha_s(0)$ and corresponding "dressing"  function are shown by solid lines in Fig.\ref{gluon_fig}.
At large Euclidean momenta the propagator (solid line in LHS of 
Fig.\ref{gluon_fig}) behaves as the free one ($1/p^2$)  and the dressing function approaches to a constant value (solid line in RHS of Fig.\ref{gluon_fig}).  In order to model the logarithmic scaling at short distances, one may divide the tree-level propagator by an appropriate logarithm\footnote{We are grateful to Jan Pawlowski who has prompted us to model the short distance scaling of the dressing function in this manner.}, which,  for background gauge,  can be attributed to the effective running coupling constant,
\begin{eqnarray}
&&\alpha_s(p) G(p)=\frac{\alpha_s(p)}{p^2}\left(1-e^{-p^2/\Lambda^2}\right), \ \ 
\alpha_s(p)=\alpha_s(0)Z(p),
\nonumber\\
&&Z(p)=\frac{12\pi}{11N_{\rm c}}\frac{1}{\ln(\zeta+ p^2/\Lambda^2)},
\
\zeta=\exp\left(\frac{12\pi}{11N_{\rm c}}\right).
\label{alphasp}
\end{eqnarray} 
Here the form of  logarithmic factor has been chosen to correspond to the form used in \cite{Maris,Fischer:2014xha,Dorkin:2014lxa}.
The result is shown by the dashed lines in Fig.\ref{gluon_fig}. One can see that the propagator itself (LHS) is not that much affected by taking into account the short distance scaling. On the contrary, the dressing function is expectedly modified at large $p^2$ (RHS). The shape of  modified by logarithmic scaling gluon correlator is in agreement with the result of  \textit{ab initio} numerical calculations presented in  \cite{Bowman,Ilgenfritz,Jan2015} and a corresponding part of \textit{ad-hoc} postulated input gluon propagator for the quark DSE  \cite{Maris,Fischer:2014xha,Dorkin:2014lxa}.
In the present set-up,  the bump in the dressing function  is due to the explicit $(1-\exp(-p^2/\Lambda^2))$ factor in  the fluctuation gluon field propagator \eqref{gprop}  in the presence of  the Abelian (anti-)self-dual  nonperturbative gluon fields. This factor is also responsible for the absence of a pole in the propagator which would correspond to the colour charged particles in the spectrum. In the infra-red limit the dressing function 
behaves as $p^2/\Lambda^2$, where  scale $\Lambda$ is exactly the same for the tree level (solid line) and UV-corrected (dashed line) dressing functions in Fig.\ref{gluon_fig}.  In our approach scale $\Lambda$, a gluon gap, is related to the scalar gluon condensate represented in terms of the Abelian (anti-)self-dual vacuum fields.  It is also notable that  identification of this gap with a kind of literally understood gluon mass would be misleading.

\begin{figure}
\begin{centering}
\includegraphics[width=0.3\textwidth]{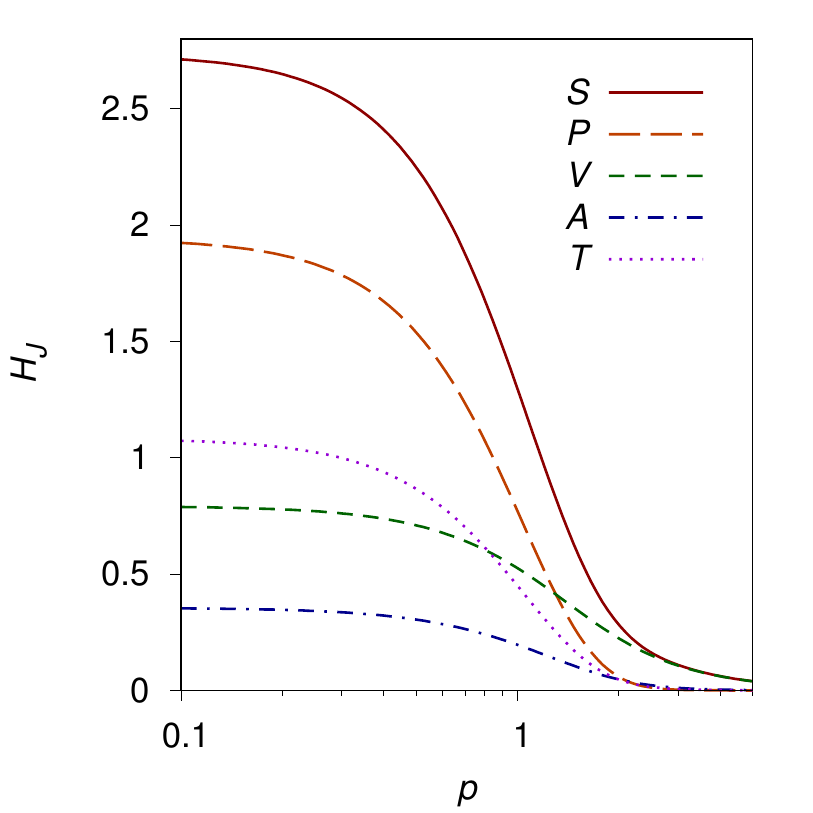}\hspace*{5mm} \includegraphics[width=0.3\textwidth]{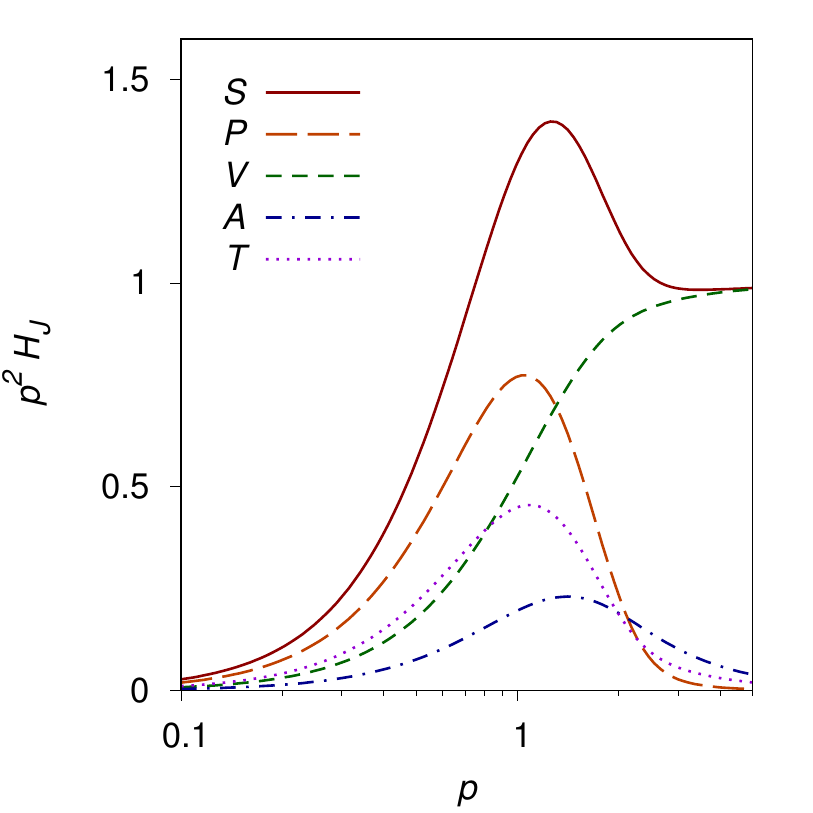}
\end{centering}
\caption{Momentum dependence of various form factors $\mathcal{H}_J$ entering the quark propagator \eqref{qpropSPVAT} and corresponding dressing functions (RHS plot) for the case of constant  quark mass ($m = 145$MeV, \ 
$\Lambda = 415$MeV) .  Dimensionless notations are used: $p^2=p^2/2v\Lambda^2$.
\label{quark_dr_SPVAT_fig}}
\end{figure}

\begin{figure}
\begin{center}
\includegraphics[scale=1]{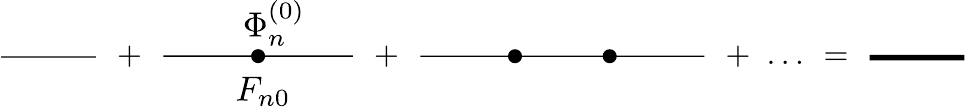}
\end{center}
\caption{Schematically represented mass corrections to the quark propagator due to the constant scalar condensates $\Phi^{(0)}_{n}$ coupled to nonlocal  form factor $F_{n0}$. Summation over the radial number $n$ is assumed.}
\end{figure}

\subsection{Quark propagator, confinement and chiral symmetry: DSE vs scaling in  the Abelian (anti-)self-dial background}

Qualitative analysis of the scaling properties can be done with regard to the quark propagator. 
The tree level propagator \eqref{qprop} takes into account the background Abelian (anti-)self-dual  field exactly and completely neglects  perturbative radiative corrections. In this approximation the quark mass has a meaning of the IR limit of the effective running mass. 

The propagator  has complicated Dirac structure \eqref{qpropSPVAT} which includes the complete set of Dirac matrices $\Gamma_J$  with corresponding form factors $\mathcal{H}_J(p^2)$ ($J=S$, $P$, $V$, $A$, $T$). For the case of constant quark mass $m(0)$ functions $\mathcal{H}_J(p^2)$ and corresponding dressing functions $p^2 \mathcal{H}_J(p^2)$ are shown in Fig.\ref{quark_dr_SPVAT_fig}. At large Euclidean momentum (short distance) the scalar and vector dressing functions tend to unity while the rest of functions vanish quickly, and the propagator approaches 
the free Dirac propagator with  mass $m(0)$.

If one switches on the momentum dependence of the quark masses and naturally assumes that $m^2(p)\ll p^2$ at large momenta $p^2\gg 1$ (we use here dimensionless notation $p^2=p^2/2v\Lambda^2$,  $m^2=m^2/2v\Lambda^2$ ) then the following simple asymptotic relations hold
\begin{eqnarray*}
&&\mathcal{H}_V=\int_0^1 ds \left(\frac{1-s}{1+s}\right)^{m^2(p)/2} e^{ -p^2 s} \xrightarrow{p^2\to\infty}\frac{1}{p^2+m^2(p)}+ O\left(e^{-p^2}\right),
\\
&&\mathcal{H}_A=\int_0^1 ds\left(\frac{1-s}{1+s}\right)^{m^2(p)/2}  s e^{ -p^2 s} \xrightarrow{p^2\to\infty}\frac{1}{(p^2+m^2(p))^2}+ O\left(e^{-p^2}\right),
\\
&&\mathcal{H}_S=\int_0^1 ds \left(\frac{1-s}{1+s}\right)^{m^2(p)/2}\frac{1}{1-s^2} e^{ -p^2 s}\xrightarrow{p^2\to\infty}\frac{1}{p^2+m^2(p)} +\frac{e^{-p^2}}{m^2(p)}+ O\left(e^{-p^2}\right),
\\
&&\mathcal{H}_P=\int_0^1 ds \left(\frac{1-s}{1+s}\right)^{m^2(p)/2} \frac{s^2}{1-s^2}e^{ -p^2 s} \xrightarrow{p^2\to\infty}\frac{2}{(p^2+m^2(p))^3}+ \frac{e^{-p^2}}{m^2(p)}
+ O\left(e^{-p^2}\right),\\
&&\mathcal{H}_T=\frac12\int_0^1 ds \left(\frac{1-s}{1+s}\right)^{m^2(p)/2}  \frac{s}{1-s^2}e^{ -p^2 s}\xrightarrow{p^2\to\infty}\frac{1}{2(p^2+m^2(p))^2} + \frac{e^{-p^2}}{2m^2(p)}+ O\left(e^{-p^2}\right).
\end{eqnarray*}
It should be stressed here that the first leading terms in above equations  have the standard perturbative form
while the subleading terms are purely nonperturbative, and they are suppressed by the Gaussian exponents.
The squared mass in the denominator of subleading terms in the last three form factors ($J=S,P,T$) originate from the  zero mode contribution to the propagator. These terms are also finite since the  scaling of the mass to zero value (in the absence of the current masses!) is suppressed by the Gaussian factor if $m^2(p)>\exp(-p^2)$ at asymptotically large $p^2$. In the limit $p^2\to \infty$ various dressing functions $p^2{\mathcal H}_J$ of the quark propagator read
\begin{eqnarray*}
p^2\mathcal{H}_V \to 1, \ p^2\mathcal{H}_S\to 1,
 \ p^2\mathcal{H}_A \to \frac{1}{p^2},
\ p^2\mathcal{H}_P\to \frac{2}{(p^2)^2}, \ 
p^2\mathcal{H}_T\to \frac{1}{2p^2} ,
\end{eqnarray*}
irrespective to the running of the quark mass.
In particular, the scalar dressing function multiplied by the mass approaches the running quark mass,
\begin{eqnarray*}
 m(p)\ p^2\mathcal{H}_S\to m(p) + \frac{e^{-p^2}}{m(p)} + O\left(m(p)e^{-p^2}\right).
\end{eqnarray*}

\begin{figure}
\begin{centering}
\includegraphics[width=0.35\textwidth]{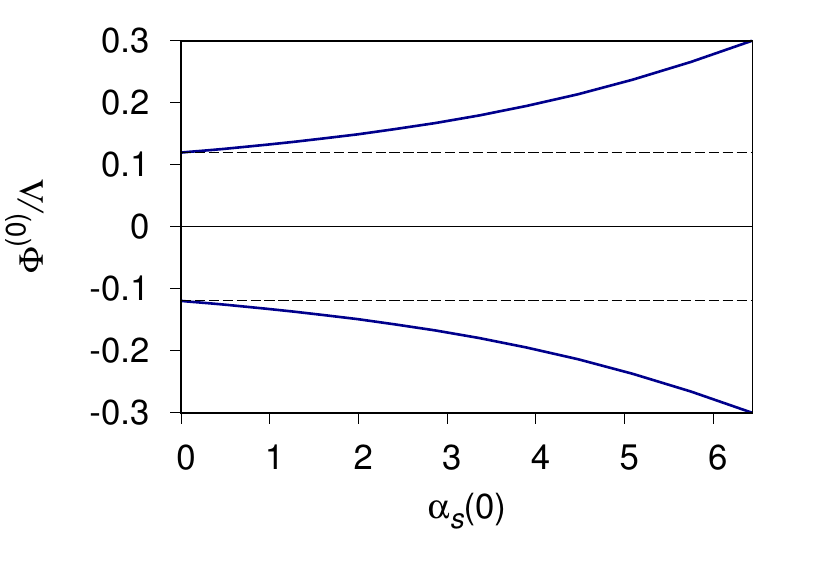}
\end{centering}
\caption{Scalar condensate $\Phi^{(0)}/\Lambda$ versus $\alpha_s(0)$ is given by solid lines, see Eq.~\eqref{Phi00}.
\label{Phi0_fig}}
\end{figure}

\begin{figure}
\begin{centering}
\includegraphics[width=0.3\textwidth]{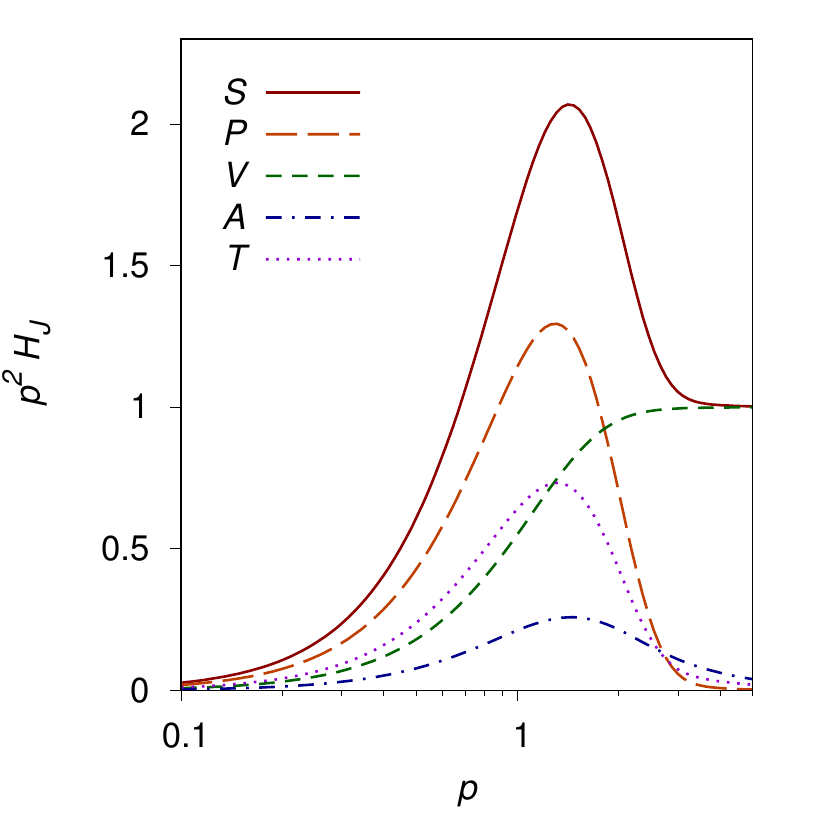}\hspace*{5mm}
\includegraphics[width=0.3\textwidth]{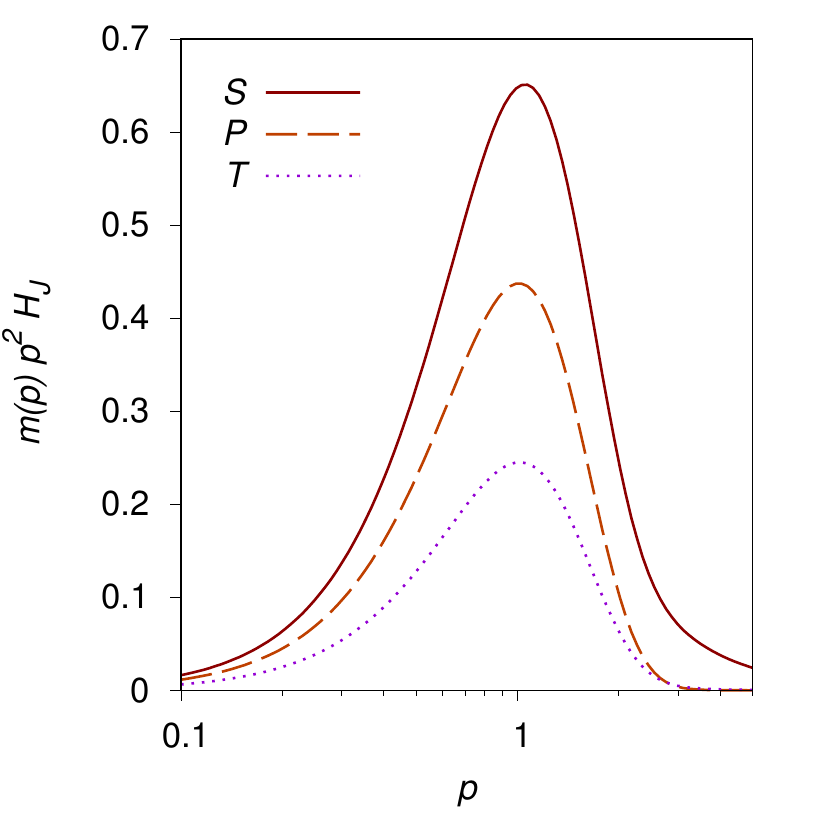}
\end{centering}
\caption{Momentum dependence of the scalar (solid line), pseudoscalar (long dash), vector (dash), axial (dash dot) and tensor (dot) dressing functions (LHS plot) in the quark propagator \eqref{qpropSPVAT} and scalar, pseudoscalar and tensor dressing functions (RHS plot) multiplied by the quark mass for the case of the running mass $m(p)$ given in Eq.\eqref{mp0}  ($m(0) = 145$MeV, \ 
$\Lambda = 415$MeV).  Dimensionless notation $p^2/2v\Lambda^2\to p^2$ is used.
\label{quark_dr_SPVAT_mp_fig}}
\end{figure}

\begin{figure}
\begin{centering}
\includegraphics[width=0.32\textwidth]{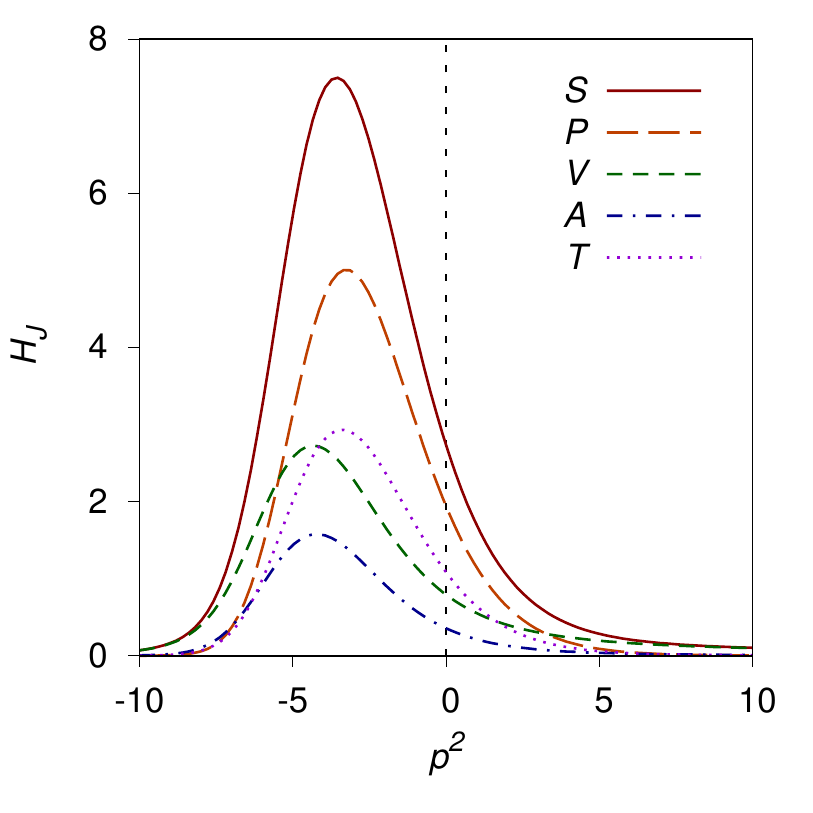}\hspace*{5mm}
\includegraphics[width=0.32\textwidth]{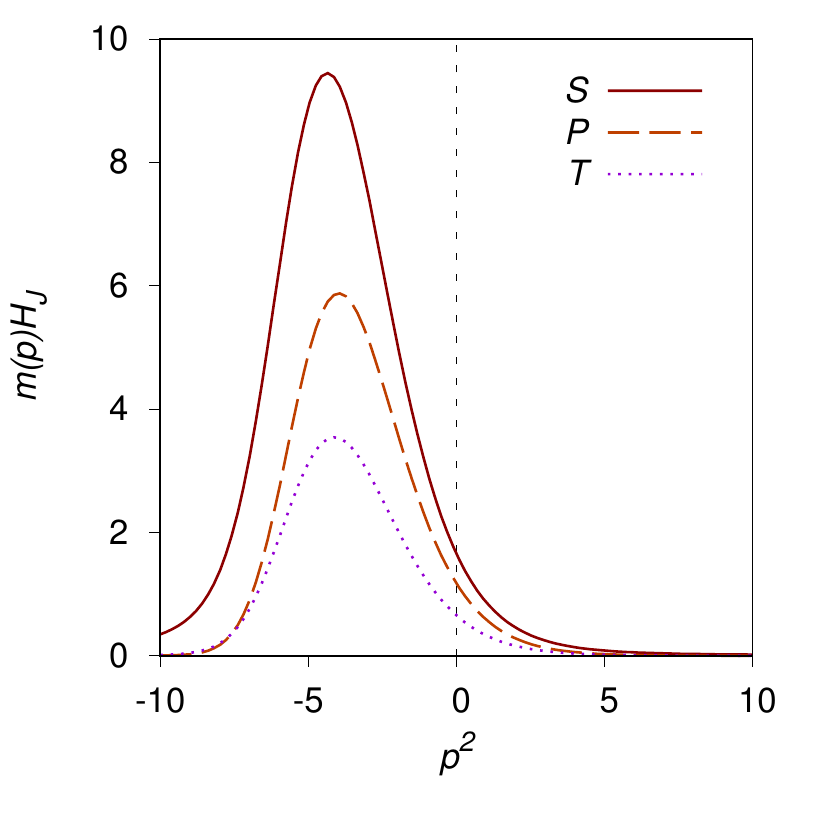}
\end{centering}
\caption{Momentum dependence of the scalar (solid line), pseudoscalar (long dash), vector (dash), axial (dash dot) and tensor (dot) form factors (LHS plot) in the quark propagator \eqref{qpropSPVAT}, and scalar, pseudoscalar and tensor form factors (RHS plot) multiplied by the quark mass  for the case of the running   mass $m(p)$ given in Eq.\eqref{mp0} with ($m(0) = 145$MeV, \ 
$\Lambda = 415$MeV). Dimensionless notation $p^2/2v\Lambda^2\to p^2$ is used.
\label{quark_SPVAT_mp_fig}}
\end{figure}

One can see that at large momenta the propagator approaches the standard Dirac propagator.
For intermediate values of momenta the dressing functions are shown in LHS of Fig.\ref{quark_dr_SPVAT_mp_fig}. The RHS of this figure emphasizes that the scalar part of the dressing function multiplied by the mass (see \eqref{qpropSPVAT}) scales at large momentum as the running mass,  while pseudoscalar and tensor structures vanish.

Equation \eqref{condensates} for scalar condensates $\Phi^{(0)}_{n0}$ suggests a verisimilar model for the  nonperturbative  running  constituent mass of the quarks 
\begin{eqnarray}
m(p)= \bar m(0)\sum_{n=0}^{\infty}C_nF_{n0}\left(-\frac{p^2}{\Lambda^2}\right),
\label{mp}
\end{eqnarray}
where we have returned to the dimensionful   notation. The coefficients $C_n$  are normalized as
\begin{eqnarray*}
\sum_{n=0}^{\infty}C_nF_{n0}\left(0\right)=1.
\end{eqnarray*}
If contributions to the quark mass corresponding to $n>0$  are neglected in Eq.\eqref{mp} then the running mass in the chiral limit
\begin{eqnarray}
m(p)= \bar m(0)F_{00}(p^2), \ \ F_{00}(p)=\left[1-\exp\left(-\frac{p^2}{\Lambda^2}\right)\right]\frac{\Lambda^2}{p^2}, \ \  \bar m(0)= \frac{1}{3}g\Phi^{(0)},
\label{mp0}
\end{eqnarray}
is defined by the equation 
\begin{equation}
\Phi^{(0)}- \frac{\alpha_s\Phi^{(0)}}{9\sqrt{2}\pi}\mathrm{Tr}_v \frac{1}{v} \int_0^\infty dp^2 p^2F^2_{00}(p)\int_0^1 ds   \left(\frac{1-s}{1+s}\right)^\frac{m^2(p)}{4v} \frac{1}{1-s^2} \exp\left(-\frac{p^2}{2v}s\right)=0,
\label{Phi00}
\end{equation}
where we have used dimensionless notation 
$$
\frac{p^2}{\Lambda^2}\to p^2, \quad \frac{m}{\Lambda}\to m, \quad \frac{\Phi_0}{\Lambda}\to \Phi_0.
$$
 As is illustrated in Fig.\ref{Phi0_fig}  this equation has two solutions  $\Phi_0(\alpha_s)$ for any $\alpha_s$.     The trivial solution $\Phi_0=0$ is absent as the integral over $s$ is singular in the limit $\Phi_0\to 0$ due to the contribution of zero modes. 
 We see that there are two contributions to the scalar condensate: the vacuum field itself and the four-fermion interaction.   Thus, due to the Abelian (anti-)self-dual  vacuum field the chiral symmetry is spontaneously broken for  arbitrarily small four-fermion interaction. 
  For the values of strong coupling constant and scale given in the Table~\ref{values_of_parameters}, $\alpha_s = 3.45$ and $\Lambda= 416$~MeV, one arrives at the estimate $\bar m(0)= 166$~MeV,  which is not that different from the value \eqref{chiral}.  
It is also very important that the quark propagator with the running mass of the form \eqref{mp} remains   an entire analytical function in the complex $p^2$ plane, and the dynamical confinement property stays intact. Moreover,  the propagator with the running mass decreases both in Euclidean and Minkowski domains of $p^2$, as it is illustrated in  Fig.\ref{quark_dr_SPVAT_mp_fig}. Similar shape of the scalar and vector form factors was reported in Dyson-Schwinger approach~\cite{Souchlas:2010boa} and interpreted in terms of  the complex conjugated poles of the quark propagator. However this is not a unique interpretation, as one may conclude from the present consideration.   

The full set of constants $\Phi_n^{(0)}$, which correspond to the invariant with respect to the local gauge transformation quark condensates $\langle\bar\psi(x) F_{n0}(\stackrel{\leftrightarrow}{\cal D}\hspace*{-0.25em}
(x)/\Lambda^2)\psi(x)\rangle$, can be obtained by means of  complete solution of Eq.\eqref{condensates} including heavy flavours. Like equation \eqref{Phi00}, in the chiral limit this system of equations is free from  divergences, both ultraviolet and infrared (for discussion of analogous quark condensates see paper~\cite{Dorokhov:2005pg}).  Complete solution of Eq.\eqref{condensates} should allow us to use  the current quark masses  as free parameters instead of the infrared limit of the nonperturbative constituent masses.

\section{Summary}
In the present approach, dependence  of the light meson masses  on radial quantum number $n$ and orbital momentum $l$  has Regge character for $n\gg 1$ or $l\gg 1$ as it has been shown long time ago~\cite{EN1}.
The source of Regge mass spectrum in the model is the same as the source of dynamical colour confinement --   the impact of the Abelian (anti-)self-dual background fields onto gluon and quark propagators resulting in the absence of singularities in propagators in the whole complex momentum plane. Gaussian exponential dependence of propagators on momentum is of particular importance for the strictly equidistant spectrum of $M_{nl}^2$ (see a highly instructive consideration of the toy nonlocal Kutkosky model in papers~\cite{EG,NKS}). 
 Nondiagonal in radial number terms in the quadratic part of the action were neglected in our previous estimations~\cite{EN,NK4}. Results of  improved computation with proper diagonalisation is given in Tables~\ref{light_mesons}, \ref{heavy-light}, \ref{heavy} and \ref{constants}.   Diagonalization significantly changes the values of parameters of the model but does not affect main features of the model.
The values of parameters given in Table \ref{values_of_parameters} were fitted to ground state mesons $\pi,\rho,K, K^*,\eta', J/\psi,\Upsilon$. The rest of masses, decay and transition constants were computed straightforwardly without further tuning of the parameters. In particular the same strong coupling constant was taken for light, heavy light mesons and heavy quarkonia. Diagonalization in combination with chiral symmetry implementation  turned out to be crucial for computation of  leptonic decay constants $f_P$ for radially excited states. In comparison with available experimental data the overall accuracy of the model is about 11-15\% (with very few exceptions like  $g_{V\gamma}$ for heavy quarkonia).

In the Discussion section we have touched on several important issues related to the chiral symmetry implementation, comparison with  FRG and DSE results for propagators, and the interrelation of the present formalism with AdS/QCD models. More systematic study of these problems will be presented elsewhere.

\section*{ACKNOWLEDGMENTS} We acknowledge fruitful discussions with
 J. Pawlowski, R. Alkofer, M. Ilgenfritz, M. Ivanov, A. Dorokhov, S. Brodsky, L. Kaptari, S. Dorkin and A. Kataev.

\appendix

\section{$U(1)$-gauging of the nonlocal meson action}
Let us start with the generating functional
\begin{multline*}
Z=\int d\sigma_\mathrm{vac} D\bar{q} Dq \exp\left\{ -\int \!\! \!\int d^4x d^4y \bar{q}_f(x) S_f^{-1}(x,y|B) q_f(y) \right.\\
\left. + g^2 \sum_{aJ}C_J \int \!\!\!\int d^4x d^4y D(y|\Lambda^2)  J^{\dagger aJ}(x,y) J^{ aJ}(x,y)\right\}.
\end{multline*}
Bilocal currents $J^{ aJ}(x,y)$ are
\begin{gather*}
J^{ aJ}(x,y)=\bar{q}_f(x+\xi y) M_{ff'}^a \Gamma^J \exp\left[-i\int_{x+\xi y}^{x-\xi' y}dz_\mu \widehat{B}_\mu(z)\right]q_{f'}(x-\xi'y)\\
\xi = \frac{m_{f'}}{m_f+m_{f'}} , \quad \xi'=\frac{m_f}{m_f+ m_{f'}}.
\end{gather*}
To make the Lagrangian gauge invariant, one performs substitution
\begin{equation*}
\partial_\mu\rightarrow \partial_\mu-ie_f A_\mu(x)
\end{equation*}
and insert the term \cite{Terning:1991yt}
\begin{equation*}
\exp\left[-ie_f\int_{x+\xi y}^x dz_\mu A_\mu(z) -ie_{f'}\int_x^{x-\xi' y}dz_\mu A_\mu(z)\right]
\end{equation*}
in $J^{ aJ}(x,y)$. Bilocal current takes the form
\begin{multline*}
J^{ aJ}(x,y|A)=\bar{q}_f(x+\xi y) M_{ff'}^a \Gamma^J\\
\exp\left[-i\int_{x+\xi y}^{x-\xi' y}dz_\mu \widehat{B}_\mu(z)-ie_f\int_{x+\xi y}^x dz_\mu A_\mu(z) -ie_{f'}\int_x^{x-\xi' y}dz_\mu A_\mu(z)\right]q_{f'}(x-\xi'y).
\end{multline*}
Expanding $J^{ aJ}(x,y|A)$ in powers of electric charge $e$, one obtains
\begin{equation*}
J^{ aJ}(x,y|A)=J^{ aJ}(x,y)
\left(1-ie_f\int_{x+\xi y}^x dz_\mu A_\mu(z) -ie_{f'}\int_x^{x-\xi' y}dz_\mu A_\mu(z)+\dots \right).
\end{equation*}
Integrals may be evaluated along the straight line:
\begin{gather*}
i\int_{x+\xi y}^{x-\xi' y}dz_\mu \widehat{B}_\mu(z) = -\frac{i}{2} x_\mu\widehat{B}_{\mu\nu}y_\nu,\\
\int_a^b dz_\mu A_\mu(z)=\int_a^b dz_\mu \int \frac{d^4p}{(2\pi)^4} \tilde{A}_\mu(p)e^{ipz}= \int \frac{d^4p}{(2\pi)^4} \tilde{A}_\mu(p) \int_a^b dz_\mu e^{ipz},\\
\int_a^b dz_\mu e^{ipz}=\int_0^1 d\tau (b-a)_\mu e^{ip(a+\tau(b-a))}=e^{ipa} \int_0^1 d\tau \frac{1}{i\tau} \frac{\partial}{\partial p_\mu} e^{ip\tau (b-a)},\\
\int_{x+\xi y}^x dz_\mu A_\mu(z) = -\int^{x+\xi y}_x dz_\mu A_\mu(z) = -\int \frac{d^4p}{(2\pi)^4} \tilde{A}_\mu(p) e^{ipx} \int_0^1 d\tau \frac{1}{i\tau} \frac{\partial}{\partial p_\mu} e^{ip\tau \xi y},\\
\int_x^{x-\xi' y} dz_\mu A_\mu(z) = \int \frac{d^4p}{(2\pi)^4} \tilde{A}_\mu(p) e^{ipx} \int_0^1 d\tau \frac{1}{i\tau} \frac{\partial}{\partial p_\mu} e^{ip\tau (-\xi' y)}.
\end{gather*}
Expansion of the bilocal current up to the first power in charge may be rewritten as
\begin{equation}\label{bilocal_current_charge_expansion}
J^{ aJ}(x,y|A)=J^{ aJ}(x,y) \left(1+\int \frac{d^4p}{(2\pi)^4} \tilde{A}_\mu(p) e^{ipx} \int_0^1 d\tau \frac{1}{i\tau} \frac{\partial}{\partial p_\mu} \left\{ie_f e^{ip\tau \xi y}-ie_{f'}e^{ip\tau (-\xi' y)}\right\}+\dots \right).
\end{equation}
Now variable $y$ may be integrated out:
\begin{equation}
\label{1-photon-vertex}
V^{aJln}_A(x)= \int_0^1 d\tau \frac{1}{\tau} \frac{\partial}{\partial p_\mu} \left\{e_f V^{aJln}\left( \stackrel{\leftrightarrow}{\nabla}(x) +ip\tau\xi\right)- e_{f'} V^{aJln}\left(\stackrel{\leftrightarrow}{\nabla}(x) - ip\tau\xi'\right)\right\}.
\end{equation}
For example, the interaction of a ground-state meson with quark current and electromagnetic field is described by the vertex
\begin{equation*}
V^{aJ00}_A=M_{ff'}^a \Gamma^J \int_0^1 d\tau \frac{1}{\tau} \frac{\partial}{\partial p_\mu} \int_0^1 dt\left\{e_f\exp\left[\frac{t}{\Lambda^2} \left( \stackrel{\leftrightarrow}{\nabla} +ip\tau\xi\right)^2 \right]- e_{f'} \exp\left[\frac{t}{\Lambda^2}\left(\stackrel{\leftrightarrow}{\cal D} - ip\tau\xi'\right)^2\right]\right\},
\end{equation*}
where $p$ is momentum of electromagnetic field.

\section{$SU(2)_\mathrm{L}\times U(1)_\mathrm{Y}$-gauging of the meson Lagrangian}
Quark fields
\begin{equation*}
Q=\left(u,c,t,d,s,b\right)^\mathrm{T}
\end{equation*}
that diagonalize mass matrix and Higgs interaction are transformed as
\begin{equation*}
Q^{\omega,\varepsilon} = 
\exp\left(
\begin{array}{cc}
ig \omega^3 T^3_u+ ig' Y_L\varepsilon & ig V \frac{\omega^1-i\omega^2}{2}\\
ig V^\dagger \frac{\omega^1+i\omega^2}{2} & ig  \omega^3 T^3_d + ig'  Y_L\varepsilon
\end{array}
\right)
Q_L+
\exp\left(
\begin{array}{ccc}
ig' Q_u \varepsilon & 0\\
0 &ig' Q_d \varepsilon
\end{array}
\right) 
Q_R
\end{equation*}
under the action of $SU(2)_\mathrm{L}\times U(1)_\mathrm{Y}$, where $V$ is CKM matrix. The following notations are employed:
\begin{gather*}
L=\frac{1-\gamma_5}{2},\quad R=\frac{1+\gamma_5}{2},\quad t^a=\frac{\sigma^a}{2}, \\
Y_L = \frac{1}{6},\quad
Y_R = \left(\begin{array}{cc}
\frac{2}{3}&0\\
0&-\frac{1}{3}
\end{array}\right)=
\left(\begin{array}{cc}
Q_u&0\\
0&Q_d
\end{array}\right),\quad
t^3=\left(\begin{array}{cc}
\frac{1}{2}&0\\
0&-\frac{1}{2}
\end{array}\right)=
\left(\begin{array}{cc}
T^3_u&0\\
0&T^3_d
\end{array}\right),\quad
t^3+Y_L= \left(\begin{array}{cc}
Q_u&0\\
0&Q_d
\end{array}\right).
\end{gather*}
To provide gauge invariance of Lagrangian, bilocal current
\begin{multline*}
J^{aJ}(x,y)=\bar{Q}_f(x+\xi y) M_{ff'}^a \Gamma^J \exp\left[-i\int_{x+\xi y}^{x-\xi' y}dz_\mu \widehat{G}_\mu(z)\right]Q_{f'}(x-\xi'y)=\\
\left(\bar{Q}_{fL}(x+\xi y) + \bar{Q}_{fR}(x+\xi y)\right) M_{ff'}^a \Gamma^J \exp\left[-i\int_{x+\xi y}^{x-\xi' y}dz_\mu \widehat{G}_\mu(z)\right]\left( Q_{f'L}(x-\xi'y) + Q_{f'R}(x-\xi'y)\right)
\end{multline*}
is modified in the following way:
\begin{gather*}
\begin{split}
J^{aJ}(x,y)\to\left\{\bar{Q}_L(x+\xi y)
P\exp\left[\int_{x+\xi y}^x dz_\mu\left(
\begin{array}{cc}
-ieQ_uA_\mu  - ig\frac{T^3_u - \sin^2\theta_W Y_L}{\cos\theta_W} Z_\mu & -i\frac{g}{\sqrt2}V W^+_\mu\\
-i\frac{g}{\sqrt2}V^\dagger W^-_\mu & -ieQ_dA_\mu  - ig\frac{T^3_d - \sin^2\theta_W Y_L}{\cos\theta_W} Z_\mu
\end{array}
\right)\right]\right.+\\
\left.\bar{Q}_R(x+\xi y) \exp\left[\int_{x+\xi y}^x dz_\mu
\left(
\begin{array}{cc}
-ieQ_u A_\mu + ig \frac{\sin^2\theta_W}{\cos \theta_W}Q_u Z_\mu & 0\\
0 & -ieQ_d A_\mu + ig \frac{\sin^2\theta_W}{\cos \theta_W}Q_d Z_\mu
\end{array}
\right)\right]\right\}_f \\
\times M_{ff'}^a \Gamma^J \exp\left[-i\int_{x+\xi y}^{x-\xi' y}dz_\mu \widehat{G}_\mu(z)\right]\times\\
\left\{ P\exp\left[\int^{x-\xi' y}_x dz_\mu\left(
\begin{array}{cc}
-ieQ_uA_\mu  - ig\frac{T^3_u - \sin^2\theta_W Y_L}{\cos\theta_W} Z_\mu & -i\frac{g}{\sqrt2}V W^+_\mu\\
-i\frac{g}{\sqrt2}V^\dagger W^-_\mu & -ieQ_dA_\mu  - ig\frac{T^3_d - \sin^2\theta_W Y_L}{\cos\theta_W} Z_\mu
\end{array}
\right)\right]Q_L(x-\xi' y)\right.+\\
\left. \exp\left[\int^{x-\xi' y}_x dz_\mu
\left(
\begin{array}{cc}
-ieQ_u A_\mu + ig \frac{\sin^2\theta_W}{\cos \theta_W}Q_u Z_\mu & 0\\
0 & -ieQ_d A_\mu + ig \frac{\sin^2\theta_W}{\cos \theta_W}Q_d Z_\mu
\end{array}
\right)\right]Q_R(x - \xi' y)\right\}_{f'},
\end{split}\\
W^{\pm}_\mu=\frac{W^1_\mu\mp iW^2_\mu}{\sqrt{2}},\quad A_\mu=\frac{g'W^3_\mu + gB_\mu}{\sqrt{g^2 + g'^2}}, \quad Z_\mu=\frac{gW^3_\mu - g'B_\mu}{\sqrt{g^2 + g'^2}},\\
e=\frac{gg'}{\sqrt{g^2 + g'^2}}, \quad \cos\theta_W = \frac{g}{\sqrt{g^2 + g'^2}}, \quad \sin\theta_W = \frac{g'}{\sqrt{g^2 + g'^2}},
\end{gather*}
where $P$ is path-antiordering (higher values of path parameter stand to the right). 

Now we are ready to investigate first-order perturbative expansion of the bilocal current. Let us consider an example of $W^+$ interaction with a charged meson. The term describing interaction of meson with quark current and $W^+$ is
\begin{multline*}
\left\{\bar{Q}_L(x+\xi y)
\int_{x+\xi y}^x dz_\mu\left(
\begin{array}{cc}
0 & -i\frac{g}{\sqrt2}V W^+_\mu\\
0 & 0
\end{array}
\right)\right\}_f  M_{ff'}^a \Gamma^J \exp\left[-i\int_{x+\xi y}^{x-\xi' y}dz_\mu \widehat{G}_\mu(z)\right]Q_{f'}(x+\xi'y)+\\
\bar{Q}(x+\xi y) M_{ff'}^a \Gamma^J \exp\left[-i\int_{x+\xi y}^{x-\xi' y}dz_\mu \widehat{G}_\mu(z)\right]\left\{ \int^{x-\xi' y}_x dz_\mu\left(
\begin{array}{cc}
0 & -i\frac{g}{\sqrt2}V W^+_\mu\\
0 & 0
\end{array}
\right)Q_L(x-\xi' y)\right\}_{f'}=\\
-i\frac{g}{\sqrt{2}}\bar{Q}_{f_1}(x+\xi y)
 \exp\left[-i\int_{x+\xi y}^{x-\xi' y}dz_\mu \widehat{G}_\mu(z)\right]\times\\
\left\{ R\Gamma^J  \left(
\begin{array}{cc}
0 & V\\
0 & 0
\end{array}
\right)_{f_1 f} M_{ff'}^a \delta_{f'f_2} \int_{x+\xi y}^x dz_\mu W^+_\mu +
\Gamma^J L \delta_{f_1f} M_{ff'}^a \left(
\begin{array}{cc}
0 & V\\
0 & 0
\end{array}
\right)_{f' f_2} \int^{x-\xi' y}_x dz_\mu W^+_\mu 
\right\} Q_{f_2}(x - \xi'y).
\end{multline*}
Proceeding in the way analogous to \eqref{bilocal_current_charge_expansion} and integrating out relative coordinate $y$, we write the desired vertex as
\begin{multline*}
V^{aJln}_{W^+f_1 f_2}= \frac{g}{\sqrt{2}} \int_0^1 d\tau \frac{1}{\tau} \frac{\partial}{\partial p_\mu}\\
\left\{R \left(
\begin{array}{cc}
0 & V\\
0 & 0
\end{array}
\right)_{f_1 f} V^{aJln}_{ff'}\left( \stackrel{\leftrightarrow}{\cal D}(x) +ip\tau\xi\right) \delta_{f'f_2} -
\delta_{f_1f} V^{aJln}_{ff'}\left( \stackrel{\leftrightarrow}{\cal D}(x) - ip\tau\xi' \right) \left(
\begin{array}{cc}
0 & V\\
0 & 0
\end{array}
\right)_{f' f_2} L\right\}.
\end{multline*}


\begin{thebibliography}{99}

 \bibitem{Feynman}R. P. Feynman, M, Kislinger, and F. Ravndal,  Phys. Rev. D {\bf 3} (1971) 2706.

\bibitem{Leutwyler:1977pv} 
  H.~Leutwyler and J.~Stern,
  Phys.\ Lett.\ B {\bf 73}, 75 (1978).

\bibitem{Leutwyler:1977vy} 
  H.~Leutwyler and J.~Stern,
  Annals Phys.\  {\bf 112}, 94 (1978).

\bibitem{Leutwyler:1977vz} 
  H.~Leutwyler and J.~Stern,
  Phys.\ Lett.\ B {\bf 69}, 207 (1977).

\bibitem{Leutwyler:1977cs} 
  H.~Leutwyler and J.~Stern,
  Nucl.\ Phys.\ B {\bf 133}, 115 (1978).

\bibitem{Leutwyler:1978uk} 
  H.~Leutwyler and J.~Stern,
  Nucl.\ Phys.\ B {\bf 157}, 327 (1979).
  
\bibitem{Karch:2006pv} 
  A.~Karch, E.~Katz, D.~T.~Son and M.~A.~Stephanov,
  Phys.\ Rev.\ D {\bf 74}, 015005 (2006)
  [hep-ph/0602229].

\bibitem{deTeramond:2005su} 
  G.~F.~de Teramond and S.~J.~Brodsky,
  Phys.\ Rev.\ Lett.\  {\bf 94}, 201601 (2005)
  [hep-th/0501022].

\bibitem{Brodsky:2006uqa} 
  S.~J.~Brodsky and G.~F.~de Teramond,
  Phys.\ Rev.\ Lett.\  {\bf 96}, 201601 (2006)
  [hep-ph/0602252].

\bibitem{deTeramond:2008ht} 
  G.~F.~de Teramond and S.~J.~Brodsky,
  Phys.\ Rev.\ Lett.\  {\bf 102}, 081601 (2009)
  [arXiv:0809.4899 [hep-ph]].

\bibitem{Swarnkar} 
  R.~Swarnkar and D.~Chakrabarti,
  Phys.\ Rev.\ D {\bf 92}, no. 7, 074023 (2015)
\bibitem{Gutsche:2015xva} 
  T.~Gutsche, V.~E.~Lyubovitskij, I.~Schmidt and A.~Vega,
  Phys.\ Rev.\ D {\bf 91},  114001 (2015).

\bibitem{EN1}   G.V. Efimov, and S.N. Nedelko,  Phys. Rev. D \textbf{51}, 176 (1995).

\bibitem{Leutwyler2} H. Leutwyler,  Nucl. Phys. B \textbf{179} , 129 (1981). 

\bibitem{Leutwyler1} H. Leutwyler,  Phys. Lett.  B \textbf{96}, 154 (1980).

\bibitem{Minkowski} P. Minkowski,  Phys. Lett. B \textbf{76},  439 (1978).

\bibitem{NV}
S.~N.~Nedelko and V.~E.~Voronin,
  Eur.\ Phys.\ J.\ A {\bf 51},  45 (2015).

\bibitem{Pagels} H. Pagels, and E. Tomboulis,  Nucl. Phys. B \textbf{143}, 485 (1978).

\bibitem{Pawlowski}
A.~Eichhorn, H.~Gies and J.~M.~Pawlowski,
Phys.\ Rev.\  D {\bf 83}, 045014 (2011).

\bibitem{NG2011} B.V. Galilo and S.N. Nedelko, Phys. Part.  Nucl.  Lett.,
\textbf{8},  67 (2011).

\bibitem{George:2012sb} 
  D.~P.~George, A.~Ram, J.~E.~Thompson and R.~R.~Volkas,
  Phys.\ Rev.\ D {\bf 87}, 105009 (2013).
  [arXiv:1203.1048 [hep-th]].

\bibitem{EN}  J.~V.~Burdanov, G.~V.~Efimov, S.~N.~Nedelko, S.~A.~Solunin,
 Phys. Rev. D\textbf{ 54}, 4483 (1996).

\bibitem{NK1}   A.C. Kalloniatis and S.N. Nedelko,  Phys. Rev. D \textbf{64}, 114025  (2001).

\bibitem{Olesen7} 
P.~Olesen, Nucl. Phys. B 200, 381 (1982).

\bibitem{NK4}   
 A.C. Kalloniatis and S.N. Nedelko,  Phys. Rev. D \textbf{69}, 074029 (2004).

\bibitem{faddeev} L.D. Faddeev, arXiv:0911.1013 [math-ph].

\bibitem{NK6} 
  A.~C.~Kalloniatis and S.~N.~Nedelko,
  Phys.\ Rev.\ D {\bf 73}, 034006 (2006).

\bibitem{roberts} 
  J.~Praschifka, C.~D.~Roberts and R.~T.~Cahill,
  Phys.\ Rev.\ D {\bf 36}, 209 (1987).

\bibitem{PDG} K.A. Olive et al. (Particle Data Group) Chinese Phys. C \textbf{38},090001 (2014).

\bibitem{Dolezal:2015gpa} 
  Z.~Dolezal [ATLAS Collaboration],
  PoS Bormio {\bf 2015}, 035 (2015).

\bibitem{Dowdall:2012ab} 
  R.~J.~Dowdall, C.~T.~H.~Davies, T.~C.~Hammant and R.~R.~Horgan,
  Phys.\ Rev.\ D {\bf 86}, 094510 (2012)
  [arXiv:1207.5149 [hep-lat]].

\bibitem{Chiu:2007bc} 
  T.~W.~Chiu {\it et al.}  [TWQCD Collaboration],
  PoS LAT {\bf 2006}, 180 (2007).

\bibitem{Holl:2004fr}
  A.~Holl, A.~Krassnigg and C.~D.~Roberts,
  Phys.\ Rev.\ C {\bf 70}, 042203 (2004).

\bibitem{krasnikov}
A.~L.~Kataev, N.~V.~Krasnikov and A.~A.~Pivovarov,
  Phys.\ Lett.\ B {\bf 123}, 93 (1983).

\bibitem{kataev}
  S.~G.~Gorishnii, A.~L.~Kataev and S.~A.~Larin,
  Phys.\ Lett.\ B {\bf 135}, 457 (1984).

\bibitem{Jan2015} M.~Mitter, J.~M.~Pawlowski and N.~Strodthoff,
  Phys.\ Rev.\ D {\bfseries 91} (2015) 054035;
  [arXiv:1411.7978 [hep-ph]].

\bibitem{Bowman}
  P.~O.~Bowman, U.~M.~Heller, D.~B.~Leinweber, M.~B.~Parappilly and A.~G.~Williams,
  Phys.\ Rev.\ D {\bfseries 70},  (2004) 034509;
  [hep-lat/0402032].

\bibitem{Ilgenfritz}
A.~Sternbeck, E.-M.~Ilgenfritz, M.~Muller-Preussker, A.~Schiller and I.~L.~Bogolubsky,
  PoS LAT {\bfseries 2006}, 076 (2006).
  [hep-lat/0610053].

\bibitem{Maris}
P. Maris and P. C. Tandy, Phys. Rev. C 60, 055214 (1999)


\bibitem{Fischer:2014xha} 
  C.~S.~Fischer, S.~Kubrak and R.~Williams,
  Eur.\ Phys.\ J.\ A {\bf 50}, 126 (2014).

\bibitem{Dorkin:2014lxa} 
  S.~M.~Dorkin, L.~P.~Kaptari and B.~Kämpfer,
  Phys.\ Rev.\ C {\bf 91} (2015) 055201,
  arXiv:1412.3345 [hep-ph].

\bibitem{Souchlas:2010boa} 
  N.~Souchlas,
  J.\ Phys.\ G {\bf 37},115001 (2010).

\bibitem{Dorokhov:2005pg} 
  A.~E.~Dorokhov,
  Eur.\ Phys.\ J.\ C {\bf 42}, 309 (2005).

\bibitem{EG} 
  G.~V.~Efimov and G.~Ganbold,
  Phys.\ Rev.\ D {\bf 65}, 054012 (2002).

\bibitem{NKS} 
  A.~C.~Kalloniatis, S.~N.~Nedelko and L.~von Smekal,
  Phys.\ Rev.\ D {\bf 70}, 094037 (2004).

\bibitem{Terning:1991yt} 
  J.~Terning,
  Phys.\ Rev.\ D {\bf 44}, no. 3, 887 (1991).



\end{thebibliography}
\end{document}